\documentclass[sigconf]{acmart}

\usepackage[utf8]{inputenc}

\usepackage{tikz}
\usepackage{amsmath}
\usepackage[nolist]{acronym}
\usepackage{enumitem}
\usepackage{rotating}
\usepackage{booktabs}
\usepackage{hyperref}
\usepackage{listings}
\usepackage{csquotes}
\usepackage{xcolor}

\newcommand{\wifi}{Wi-Fi}
\newcommand{\smarthome}{Smart Home}
\newcommand{\zigbee}{ZigBee}
\newcommand{\bluetooth}{Bluetooth}

\newcommand{\nmap}{NMAP}
\newcommand{\OSC}{Open Source Code}
\newcommand{\telnet}{Telnet}
\newcommand{\sslpinning}{SSL Pinning}

\newcommand{\arlocam}{Arlo Essential Indoor Security Camera}
\newcommand{\blinkcam}{Blink Mini Camera}
\newcommand{\boschcam}{Bosch 360\textdegree{} Indoor Camera}
\newcommand{\dlinkcam}{D-Link DCS-8010LH Camera}
\newcommand{\nestcam}{Nest Indoor Camera}
\newcommand{\nestprotect}{Nest Protect Smoke \& CO Alarm Detector}
\newcommand{\ringcam}{Ring Spotlight Camera}
\newcommand{\tesvor}{Tesvor S6 Vacuum Cleaner}
\newcommand{\tpcam}{TP-Link Tapo C100 Camera}

\newenvironment{compactitemt}{\begin{itemize}[nosep, wide=0pt,
  labelsep=1pt,
  before={\begin{minipage}[t]{\hsize}},
  after ={\end{minipage}} ]   }{ \end{itemize} }
\newenvironment{smallitemize}{\begin{compactitemt}}{\end{compactitemt}}

\def\twodigits#1{%
  \ifnum#1<10 0\fi
  \number#1}

\usepackage{microtype}
\usepackage{textcase}

\DeclareRobustCommand{\spacedlowsmallcaps}[1]{%
  \textls[80]{\scshape\MakeTextLowercase{#1}}%
}
 
\usepackage{tabularx} 
\newcommand{\tableheadline}[1]{\multicolumn{1}{l}{\spacedlowsmallcaps{#1}}}
\newcommand{\myfloatalign}{\centering} 

\newcommand{\newtag}[2]{#1\def\@currentlabel{#1}\label{#2}}
\newcommand{\newshorttag}[2]{\def\@currentlabel{#1}\label{#2}}

\setcopyright{none}
\acmConference[To appear in ACSAC 2023]{ACM Conference}{December 2023}{Austin, Texas, USA}

\begin{document}

\begin{acronym}[UMLX]
	\acro{AC}{Access Control}
	\acro{AE}{Authentication}
	\acro{AO}{Authorization}
	\acro{API}{Application Programming Interface}
	\acro{ARP}{Address Resolution Protocol}
	\acro{BLE}{Bluetooth Low Energy}   
	\acro{BP}{Best Practice}
	\acro{CRYPTO}{Cryptography}
	\acro{CSRF}{Cross-Site Request Forgery}
	\acro{CVE}{Common Vulnerabilities and Exposures}
	\acro{CVSS}{Common Vulnerability Scoring System}
	\acro{DDNS}{Dynamic Domain Name System}
	\acro{DDoS}{Distributed Denial-of-Service}
	\acro{DNS}{Domain Name System}
	\acro{DoS}{Denial-of-Service}
	\acro{DPC}{Data Protection and Compliance}
	\acro{DRY}{Don't Repeat Yourself}
	\acro{DSP}{Digital Signal Processor}
	\acro{DSP}{Strong Default Security and Privacy}
	\acro{EMC}{Electro Magnetic Compatibility}
	\acro{ENISA}{European Union Agency For Network And Information Security}
	\acro{FU}{Secure Software / Firmware Updates}
	\acro{HKDF}{HMAC-based Key Derivation Function}
	\acro{HSTS}{HTTP Strict Transport Security}
	\acro{HW}{Hardware Security}
	\acro{ICMP}{Internet Control Message Protocol}     
	\acro{IG}{Information Gathering}
	\acro{IoT}{Internet of Things}
	\acro{JWT}{JSON Web Token}
	\acro{LOG}{Logging}
	\acro{MAT}{Monitoring, Auditing and Testing} 
	\acro{MatE}{Man-at-the-End}
	\acro{MCU}{Microcontroller}
	\acro{MitM}{Man-in-the-Middle}
	\acro{NA}{Nearby Attacker}
	\acro{ONVIF}{Open Network Video Interface Forum}
	\acro{OTA}{Over-The-Air}
	\acro{PA}{Physical Attacker}
	\acro{PCB}{Printed Circuit Board}
	\acro{RCP}{Remote Control Protocol}
	\acro{RTOS}{Real-Time Operating System}     
	\acro{RTSP}{Real-Time Streaming Protocol}
	\acro{SA}{Shell Access}
	\acro{SIOH}{Secure Input and Output Handling}
	\acro{SIS}{Secure Interfaces and Network Services}
	\acro{SNA}{Same Network Attacker}
	\acro{SoC}{System on a Chip}
	\acro{SPI}{Serial Peripheral Interface}
	\acro{SSR}{System Safety and Reliability}
	\acro{SSRF}{Server-Side Request Forgery}
	\acro{STC}{Secure and Trusted Communications and Operating System}
	\acro{TC}{Test Case}
	\acro{TIM}{Trust and Integrity Management}
	\acro{UART}{Universal Asynchronous Receiver Transmitter}
	\acro{UML}{Unified Modeling Language}	
\end{acronym}

\date{}

\title{Unleashing IoT Security: Assessing the Effectiveness of Best Practices in Protecting Against Threats}

\author{Philipp Pütz}
\email{philipp-puetz@t-online.de}
\affiliation{%
  \institution{Technical University of Darmstadt}
  \country{Germany}
}

\author{Richard Mitev}
\email{richard.mitev@trust.tu-darmstadt.de}
\affiliation{%
  \institution{Technical University of Darmstadt}
  \country{Germany}
}

\author{Markus Miettinen}
\email{markus.miettinen@trust.tu-darmstadt.de}
\affiliation{%
  \institution{Technical University of Darmstadt}
  \country{Germany}
}

\author{Ahmad-Reza Sadeghi}
\email{ahmad.sadeghi@trust.tu-darmstadt.de}
\affiliation{%
  \institution{Technical University of Darmstadt}
  \country{Germany}
}


\begin{abstract}
The Internet of Things (IoT) market is rapidly growing and is expected to double from 2020 to 2025. The increasing use of IoT devices, particularly in smart homes, raises crucial concerns about user privacy and security as these devices often handle sensitive and critical information. Inadequate security designs and implementations by IoT vendors can lead to significant vulnerabilities.

To address these IoT device vulnerabilities, institutions, and organizations have published IoT security best practices (BPs) to guide manufacturers in ensuring the security of their products. However, there is currently no standardized approach for evaluating the effectiveness of individual BP recommendations. This leads to manufacturers investing effort in implementing less effective BPs while potentially neglecting measures with greater impact.

In this paper, we propose a methodology for evaluating the security impact of IoT BPs and ranking them based on their effectiveness in protecting against security threats. Our approach involves translating identified BPs into concrete test cases that can be applied to real-world IoT devices to assess their effectiveness in mitigating vulnerabilities. We applied this methodology to evaluate the security impact of nine commodity IoT products, discovering 18 vulnerabilities. By empirically assessing the actual impact of BPs on device security, IoT designers and implementers can prioritize their security investments more effectively, improving security outcomes and optimizing limited security budgets.
\end{abstract}

\maketitle

\section{Introduction}

The \ac{IoT} market reached a significant milestone, with a value of \$330.6 billion in 2020. Prognoses expect a continuous growth trajectory, reaching \$875.0 billion by 2025 \cite{MDF2021}. However, along with this growth comes the challenge of ensuring the security of IoT and innovative home products against malicious attacks. The increasing importance of these devices in our daily lives makes them attractive targets for cybercriminals in private and industrial settings.

Severe attacks such as the \textit{Mirai} botnet \cite{} and alike have infected large numbers of devices and caused widespread disruption through large-scale \ac{DDoS} attacks. These incidents have prompted IoT manufacturers to prioritize product security measures\cite{NJCCIC2016, Dange2019}. This increasing awareness has led to a surge in IoT security spending, with investments rising from around \$240 million in 2016 to \$631 million in 2021 \cite{Gartner2018, Dataprot2022}. Projections indicate that this trend will continue, with spending expected to reach \$6 billion by 2023 \cite{Juniper2020}.

Reliable and accurate measurement of the \emph{security gains}  associated with particular investments into security measures is a crucial requirement for being able to steer these investments in the right direction while at the same time using available resources as efficiently as possible. Unfortunately, measuring such security gains is not trivial, as a lack of conclusive security metrics makes assessing the overall achievable security level hard. Related works have proposed risk assessment approaches~\cite{bonilla2017metric,leversage2008estimating,mcqueen2006time,tupper2008vea,lai2007using}. However, they do not provide clear guidance for implementers on which security aspects to prioritize in practice. Proposed test suites and frameworks \cite{Tekeoglu2016, Loi2017, Costa2019, Alharbi2018} fall short of completeness. Therefore, we aggragate a suite of test cases from and based on \ac{IoT} security \aclp{BP} addressing the full range of IoT-related security issues and seek to provide a framework guiding practitioners in how to prioritize investment in security measures to maximally reduce potential security harm to their products.

To evaluate the effectiveness of security measures, one has to evaluate the \emph{security outcomes} that can be observed when applying particular security measures. 
One way to evaluate security outcomes is to look at the \emph{harm} experienced with and without applying the security measures. 
Their effectiveness can thus be measured as the degree to which harm can be reduced when applying the measures.
If the experienced harm decreases, the applied security measures have been successful.
Harm can be measured through real-world reports about incidents and the cost of damages caused by them or by estimating the potential severity of vulnerabilities discovered in real-world devices. We adopt the latter approach.

Based on these considerations, we thus seek to provide a methodology for \emph{empirically evaluating} the effectiveness of \ac{IoT} best practice security measures based on the reduction of potential harm that can be observed in real-world \ac{IoT} devices. The results of such evaluations can then be used in making informed decisions on prioritizing individual \aclp{BP} and thus optimize investment in \ac{IoT} security spending.

To achieve this goal, we first collect a large body of \ac{IoT} security \aclp{BP} from various sources: foundations \cite{ISF2019}, communities \cite{IEEECommunity2017}, governmental institutions \cite{EUAFNAIS2017, CultureUKD2018}, research papers \cite{Payne2017} and major \ac{IoT} cloud providers \cite{Sharma2019}. The collected \aclp{BP} are then used to derive a set of test cases that will then be used to analyze the security of a number of \smarthome{} Devices, providing a basis for the evaluation of the potential harm that these \smarthome{} Devices are vulnerable to. This allows us then to empirically evaluate the efficacy of individual \aclp{BP} so that they can be ranked and prioritized with regard to their effectiveness.

Our contributions can be summarized as follows:
\begin{enumerate}
	\item We introduce a methodology for empirically evaluating the efficacy of \ac{IoT} \aclp{BP} based on estimating the reduction in potential harm that individual \aclp{BP} affect (cf. \autoref{sect:methodology}).
        \item Our approach estimates the reduction of potential harm by evaluating the impact of \aclp{BP} on observable device vulnerabilities and quantitatively measuring the CVSS scores associated with them for prioritizing or ranking the \aclp{BP} (cf. \autoref{sec:experiment}).
        \item Our analysis is based on an extensive analysis of recent \ac{IoT} \aclp{BP} proposed by numerous relevant players in the \ac{IoT} security community and use these as a basis for a comprehensive collection of \ac{IoT} \aclp{BP} (cf. \autoref{sec:bestpractices}).
        \item We empirically evaluate the \aclp{BP} on a number of representative real-world \ac{IoT} devices and use the results and base our ranking ranking \aclp{BP} on the analysis results (cf. \autoref{sect:evaluation}). 
\end{enumerate}

\section{Problem Setting and Adversary Model}
Due to the proliferation of \ac{IoT} more and more attacks are targeting \ac{IoT} devices, leading to a number of devastating attacks causing potentially large-scale damages (e.g., Mirai~\cite{antonakakis2017understanding}). Many of the attacks were caused by vulnerabilities in IoT devices due to immature security designs and implementation practices of IoT hardware vendors focusing more on time-to-market than resilient and robust security solutions for their devices. 

\subsection{IoT Security Vulnerabilities}
Recently published attacks can be classified into following broad categories: \emph{Root Access Vulnerabilities}, allowing an attacker to gain full access to devices~\cite{Paleari2011, Heffner2013, Crowley2013, Calmejane2013, Tekeoglu2015}, \emph{Information Disclosure Attacks}, enabling an attacker to gain insights about the target in order to enable potential subsequent advanced attacks~\cite{Favaretto2019, Abdalla2020}, \emph{Local Onboarding Network-based Attacks}, enabling attackers to exploit the initial onboarding procedure~\cite{Bitdefender2019}, 
\emph{Hardware Attacks} utilizing physical tampering of the device to stage the attack~\cite{Wardle2014, Hernandez2014, Michele2014}, and, \emph{\ac{IoT} (State) Identification Attacks} extracting information about the state or status of devices~\cite{Copos2016, Apthorpe2017, Acar2020}. We discuss these attacks in more detail in \autoref{sec:related}.

\subsection{IoT Security Best Practices}
As the IoT market matures, more attention is being given to the secure implementations of IoT devices. To encounter the above-mentioned attack types, a number of organizations and players in the \ac{IoT} security community have defined a number of \aclp{BP} (also called Good Practices or Security Guidelines) with the goal of encountering attacks by removing vulnerabilities related to most common security threats in IoT designs and implementations enabling them. Examples of such \aclp{BP} focusing on different aspects of \ac{IoT} devices include, e.g., \textit{Secure Design Best Practices Guides}~\cite{ISF2019} or the \textit{Baseline Security Recommendations for IoT}~\cite{EUAFNAIS2017}.

However, while it is clear that \aclp{BP} can indeed be used to improve IoT device security, there is a lack of understanding of the impact of individual measures on the practical security of IoT devices. We, therefore, develop a methodology for measuring the impact of security measures like security \aclp{BP} on the observed security of IoT devices.

\subsection{Adversary Model}\label{sec:adv}

\begin{figure}
    \centering
    \includegraphics[width=0.95\linewidth, trim={8.8cm 9.0cm 8.7cm 2.5cm},clip]{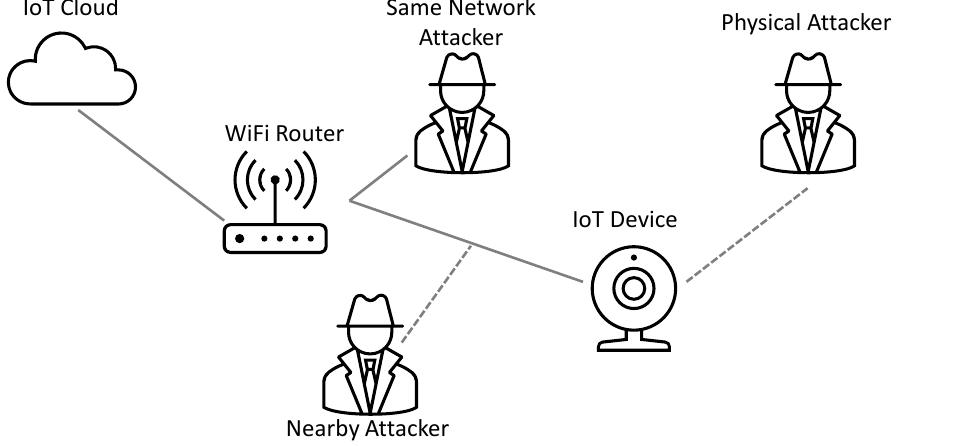}
    \caption{Adversary model in a \smarthome{} setting}
    \label{fig:attackers}
    \vspace{-1em}
\end{figure}

In the scope of this work, we focus on \aclp{BP} targeting the implementations of IoT devices and their associated communication channels. Possible broader connected ecosystems, e.g., backend services and apps, are outside the scope of this paper.

We model the adversary in terms of three distinct attacker types as shown in \autoref{fig:attackers} based on the attacker's specific connectivity and locality constraints.

The \textit{\acf{NA}} is an adversary that is located within \wifi{} or \bluetooth{} range of the victim \smarthome{} device. The main attack vector of this adversary is related to the onboarding process during the installation of the device, when the device has not yet joined the victim Smart Home network and its communication interface is thus still accessible also to external entities.

The \textit{\acf{SNA}} is an adversary that is located on the same local network as the victim device (the adversary could be, e.g., a compromised device in the victim's network).
The attacker can target traffic from the victim device to the \ac{IoT} cloud as a \ac{MitM} using \ac{DNS} or \ac{ARP} spoofing~\cite{Rapid7}.
Additionally, the \ac{SNA} can target locally exposed administration interfaces like web pages or \telnet{} shells on the victim device, which may allow root access through command injection vulnerabilities or weak passwords used to secure the \telnet{} interface.

Note that in order to simplify our adversary model, we do not consider \textit{remote attackers}, i.e., network attackers outside the local network separately, as the capabilities of a remote attacker represent a subset of what a \ac{SNA} can perform.

The \textit{\acf{PA}} is someone who has physical access to the device but is not allowed to do so, in contrast to the \textit{authorized \ac{MatE}}.
This attacker can target wireless connections like the nearby attacker and, in addition, physical ones, including the device's hardware itself.
This adversary will use, e.g., unprotected USB ports, debug interfaces, or removable storage devices to infiltrate the victim device and attempt to gain administrative privileges on the device.

\section{Methodology}
\label{sect:methodology}

\begin{figure*}
    \centering
    \includegraphics[width=0.95\textwidth, trim={1cm 6.0cm 3.5cm 7.0cm},clip]{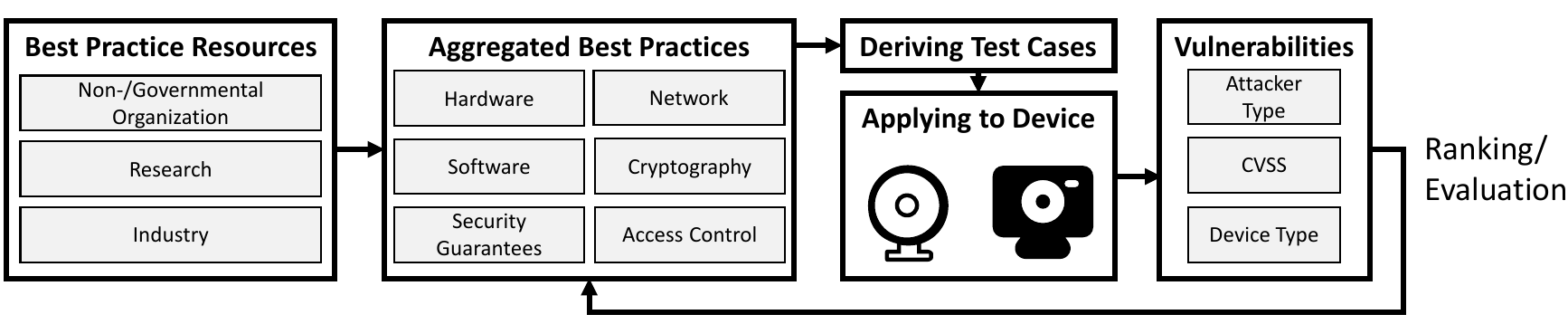}
    \caption{Overview of the pipeline of our methodology}
    \label{fig:overview}
\end{figure*}

Our methodology addresses a core challenge: how to measure the impact of individual security measures mandated by \aclp{BP}.
Ideally, one would want to measure the security impact directly as a fulfillment degree of the (ideal) goal.
\subsection{High-Level Approach}

 Our approach consists of distinct steps as shown in \autoref{fig:overview}: First, we aggregate a body of \aclp{BP}, from which we derive IoT test cases evaluated on real-world IoT devices. As a result of the evaluation, we can observe to what degree individual \aclp{BP} can reduce the potential harm by eliminating threats against the tested devices. In the following, we shall discuss the steps in more detail.

We start by first analyzing different \ac{IoT} security \aclp{BP} and recommendations and collect a comprehensive set of \ac{IoT} \aclp{BP} provided by various organizations, researchers, and industry players. Since not all \aclp{BP} can be directly applied to \smarthome{} \ac{IoT} devices, we filter them to include only such that are relevant to the development and deployment processes of \ac{IoT} devices. As a result, we obtain an aggregated collection of \aclp{BP} addressing various aspects of \ac{IoT} devices' security design, as discussed in detail in \autoref{sec:bestpractices}.

To test the effectiveness of individual \aclp{BP}, a set of test cases is derived from the aggregated set of \aclp{BP}. These test cases are evaluated on representative \smarthome{} IoT devices.
A challenge here is that the understanding of what are \enquote{\aclp{BP}} differs surprisingly much in literature, and up to 91\% of all \aclp{BP} do not describe practical actions, but desired outcomes~\cite{Bellman2020}.
Therefore, we first analyze different \ac{IoT} security \aclp{BP}, then define a common set of practices, and finally derive test cases to cover the selected \aclp{BP}.
Each test case is assigned to one or more types of attackers to facilitate testing since not all test cases may be relevant or feasible for all our attacker types.
However, since each attacker type and the corresponding test cases focus on different aspects of the device under investigation, we can maximize the attack surface to be covered.
The derived test cases are described in detail in \autoref{sec:TestCaseSelection}. 

Finally, the outcomes of the test cases are evaluated based on a threat assessment scoring system like \ac{CVSS}. This provides a quantitative estimate of the impact of individual \aclp{BP} on the threats in concrete devices and can subsequently be used to prioritize or rank individual \aclp{BP}.

\subsection{IoT Security Best Practices}
        \label{sec:bestpractices}
  
  The \ac{ENISA} lists in total 15 subcategories of Technical Measures and 9 Policies and Organizational Measures in their \textit{Baseline security measures for IoT}~\cite{EUAFNAIS2017}.
		Similarly, other guidelines like the \textit{Secure Design Best Practices Guides}~\cite{ISF2019}, the \textit{Internet of Things (IoT) Security Best Practices}~\cite{IEEECommunity2017}, the \textit{Security \aclp{BP} for the Internet of Things with Google Cloud}~\cite{Sharma2019} and the \textit{Code of Practice for Consumer IoT Security}~\cite{CultureUKD2018} also mainly describe technical recommendations for \ac{IoT} vendors and cloud providers to secure devices.
		In contrast to these, the \textit{Best Practices for Deploying IoT Devices}~\cite{Payne2017} guideline mainly describes technical measures that end customers can apply.
		Technical Measures seem to be, therefore, the most important recommendations which should be followed.
		These measures can also be studied based on available public information and do not rely on vendor-specific internal knowledge.
		Therefore, we focus on technical measures categorized according to the finer 15 subcategories from \ac{ENISA}~\cite{EUAFNAIS2017} and, where applicable, enrich them with additional measures from the above-mentioned additional guidelines.
		This common set of Technical \aclp{BP} forms the basis for the subsequent security analysis, structured by attacker types.
  	In the following, we provide an overview of the topics that applicable \aclp{BP} cover. For a detailed description of the aggregated \aclp{BP}, please refer to \autoref{sec:app1}.
	
	\paragraph{\acl{HW}}
			Hardware security uses physical elements such as chips, processors, and special \ac{PCB} designs to prevent hardware- and software-based data manipulation or access to security-relevant keys and code.
\vspace{-0.7em}
		\paragraph{\acf{TIM}}
            The main objective is to establish a verified chain of trust from the device's power-up to every executed code fragment.
\vspace{-0.7em}
		\paragraph{\acf{DSP}}
			Easy-to-guess default passwords often lead to many undesirable security issues and must therefore be avoided.
\vspace{-0.7em}
		\paragraph{\acf{DPC}}
			Emerging new types of \ac{IoT} Devices often collect sensitive information about their users, making data protection and compliance with privacy regulations more important than ever.
\vspace{-0.7em}
		\paragraph{\acf{SSR}}
			\ac{IoT} systems should return to a safe state in the event of a failure and not rely on connected services to function properly.
\vspace{-0.7em}
		\paragraph{\acf{FU}}
			During the lifetime of an \ac{IoT} Device, the discovery of vulnerabilities is very likely. Therefore, it is crucial to implement a firmware update mechanism to address vulnerabilities.
\vspace{-0.7em}
		\paragraph{\acf{AE}}
			System services and login mechanisms should use a strong authentication scheme that does not use default credentials in production and incorporates two-factor authentication.
\vspace{-1.7em}
		\paragraph{\acf{AO}}
			All users and system services should be assigned the absolute minimum privileges needed.
\vspace{-0.7em}
		\paragraph{\acf{AC}}
			Physical access control focuses on protecting the device and the data from unauthorized access.
\vspace{-0.7em}
		\paragraph{\acf{CRYPTO}}
			Strong cryptographic algorithms should secure sensitive data and connections to other services.
\vspace{-0.7em}
		\paragraph{\acf{STC}}
			Measures addressing secure and trusted communication focus on encrypting data in transmission to other services and verifying communication partners before trusting them.
\vspace{-0.7em}
		\paragraph{\acf{SIS}}
			Network interfaces and services are in many cases an essential part of a systems security posture. Especially insecure interfaces connected to the internet make large-scale attacks against \ac{IoT} devices easy.
\vspace{-0.7em}
		\paragraph{\acf{SIOH}}
			In addition to network services, administrative web pages or login forms can be used to gain unauthorized access to the \ac{IoT} device via unfiltered inputs.
\vspace{-0.7em}
		\paragraph{\acf{LOG}}
			To detect and mitigate security-related events it is important to log, e.g., failed login attempts.
\vspace{-0.7em}
		\paragraph{\acf{MAT}}
			Exploitable vulnerabilities can emerge unexpectedly at any time. Therefore, it is necessary to monitor the integrity of a system and perform \ac{IoT} security measure reviews on a regular basis to adapt them to new threats.

	\subsection{Best Practice Selection and Test Case Definition}\label{sec:TestCaseSelection}

		Some of the collected \aclp{BP} cannot be studied using publicly available information (e.g., details of proprietary closed-source implementations) or concern aspects that are not directly related to \ac{IoT} device security. Therefore, a selection of applicable \aclp{BP} has to be made.
		Additionally, as mentioned earlier, some \aclp{BP} merely describe desired outcomes and are not sufficiently precise.
		We, therefore, refine all selected \aclp{BP} into actionable test cases that can act as the basis for subsequent evaluation on real IoT devices.

		For each subcategory of the technical measures discussed above, we define different test cases that cover the scope of the selected \aclp{BP}. The detailed mapping can be found in \autoref{sec:app3}.
		If, for some reason, a \acl{BP} can not be tested, we additionally indicate the reason for this is \autoref{sec:bps}.

\section{Experimental Setup}\label{sec:experiment}
		To align the test cases (\autoref{sec:app3}) with our attacker model presented in \autoref{fig:attackers}, we associate each attacker type with one or more \textit{attack scenarios} shown in \autoref{fig:relation} that focus on different attack surfaces of the \smarthome{} Device. We then enumerate for each attack scenario all those test cases that are applicable to it (cf. \autoref{tab:scenario}).

		The testing scenario with three attackers types and a device under test is shown in \autoref{fig:attackers}.
		A router connects the setup with the internet and \ac{IoT} clouds, providing \wifi{} and Ethernet connectivity.
		A connected desktop computer is used to simulate the \acl{SNA} and network analysis tools such as \nmap{}\footnote{\url{https://nmap.org/}} or \ac{MitM} software are used on this computer.
		To facilitate network traffic analysis, the router uses port mirroring to duplicate all network traffic sent and received by the \wifi{} access point.

		The \wifi{} access point is used to connect each \smarthome{} device and a smartphone that has the respective device-specific \smarthome{} Apps installed.
        A \ac{DNS} Server is used to log \ac{DNS} requests and manipulate \ac{DNS} records to simulate \ac{DNS} Spoofing.
		Furthermore, there is a \acl{NA} within \wifi{} and \bluetooth{} range of the \smarthome{} Device and smartphone during the onboarding process.
		In this test setup, the \acl{NA} focuses only on direct connections to the \smarthome{} Device, because the \wifi{} network from the access point is assumed to be secured with a strong WPA2 password.
		Moreover, the \acl{SNA} can execute all attacks of the \acl{NA} and thus simulate an unencrypted local \wifi{} network.
		The physical attacker focuses only on hardware attacks and has direct access to ports and the encasing of the \smarthome{} Device.

    	The test cases for all applicable test scenarios are executed for each \smarthome{} Device.
    	Based on the findings of the test results a \ac{CVSS} base score is calculated for each discovered vulnerability to enable classification and ranking of discovered security issues.
    	The \ac{CVSS} base score considers aspects such as breach of confidentiality, alteration of integrity or availability, and the complexity of an attack to calculate a score between 0 (no severity) and 10 (critical severity) for a vulnerability~\cite{FIRST}.

		In the following, we discuss the test cases for each attacker type. Each device analysis starts with a general information-gathering phase, followed by execution of the attacker-specific test cases, and concludes with a batch of test cases related to shell access.
		This is because shell access test cases can only be executed in case one of the preceding test cases are successful in gaining (root) shell access to the device. The purpose of the shell access test cases is to investigate the overall security of the operating system of the tested device and installed software.

		\subsection{\acl{IG} Phase}
            \label{sec:information_gathering}

			The goal of the information-gathering phase is to collect 
			available public information about the analyzed \smarthome{} Device and its environment.			
			This includes the smartphone app, firmware version, and published \OSC{}. 
			The information gathering phase is the basis for all types of attackers and includes the steps described in \autoref{tab:IGSteps}.

			\begin{table}[t]
				\myfloatalign
				\caption{\acf{IG} Steps}
				\label{tab:IGSteps}
    \begin{scriptsize}
				\begin{tabularx}{\linewidth}{lX}
					\toprule
					\tableheadline{Step} & \tableheadline{Description}
					\\ \midrule
					01                & Determine the version of the smartphone app and if the app can be decompiled and whether obfuscation has been applied. \\
					02                & Determine the latest firmware version of the \smarthome{} Device.                                                      \\
					03                & Check if the device is cloud-only or also provides local access.                                                       \\
					04                & Search for published \OSC{} and check whether it contains hardcoded credentials.                                       \\
					05                & Search for downloadable firmware updates.                                                                              \\
					\bottomrule
				\end{tabularx}
    \end{scriptsize}
    \vspace{-1em}
			\end{table}

		\subsection{\acl{NA} Test Cases}\label{sec:TestSuite:Nearby}

			The \acl{NA} focuses on the communication between the smartphone app and the \smarthome{} Device during the onboarding process and therefore takes a look at the smartphone app or wireless connections, protocols, and services offered by the device for this purpose.
			The attack scenarios in \autoref{tab:scenario} for the \acl{NA} (1 \& 2) include test cases from the \nameref{sec:TestCaseSelection:DefaultSecurity}, \nameref{sec:TestCaseSelection:Authentication} and \nameref{sec:TestCaseSelection:Communication}, \nameref{sec:TestCaseSelection:Interfaces} and \nameref{sec:TestCaseSelection:InputOutput} categories in \autoref{sec:app3}.

			Test Cases are executed by gathering information about the services in use and searching for existing exploits, \nmap{} scanning and utilizing of sslyze\footnote{\url{https://github.com/nabla-c0d3/sslyze}} to retrieve SSL information from API services.
			In case of an open \wifi{} access point being used for the onboarding process  a \wifi{} network card in promiscuous mode is used together with Wireshark\footnote{\url{https://www.wireshark.org}} to capture the traffic (from \autoref{tab:scenario} (1)).
			Ideally, the onboarding network uses a strong WPA2 or WPA3 encryption or a secure \bluetooth{} pairing method rather than the \textit{Just Works} method.
			If that is not the case, for an unencrypted onboarding \wifi{} access point and TLS-protected HTTP traffic, Burp Suite\footnote{\url{https://portswigger.net/burp}} is used as a \ac{MitM} proxy.
			\bluetooth{} traffic is captured directly via the Android smartphone's Host Controller Interface Snoop Mode to avoid more complicated setups with external equipment.			

			User-changeable passwords (\autoref{tab:scenario} (2)) are considered as secure if they must be at least eight characters long and have a high level of complexity, i.e., are extremely long or contain different types of characters such as numbers, letters and symbols as recommended by Grassi et al.~\cite{Grassi2017}.
			Brute-force protection is considered a successful measure if the requests are slowed down, i.e., only a few requests per second are possible or are completely blocked.
			If it is possible to communicate with the onboarding services using a device other than the smartphone app, this is considered a security issue.

\begin{figure}
    \centering
    \includegraphics[width=0.95\linewidth, trim={7.5cm 5.4cm 13.3cm 2.4cm},clip]{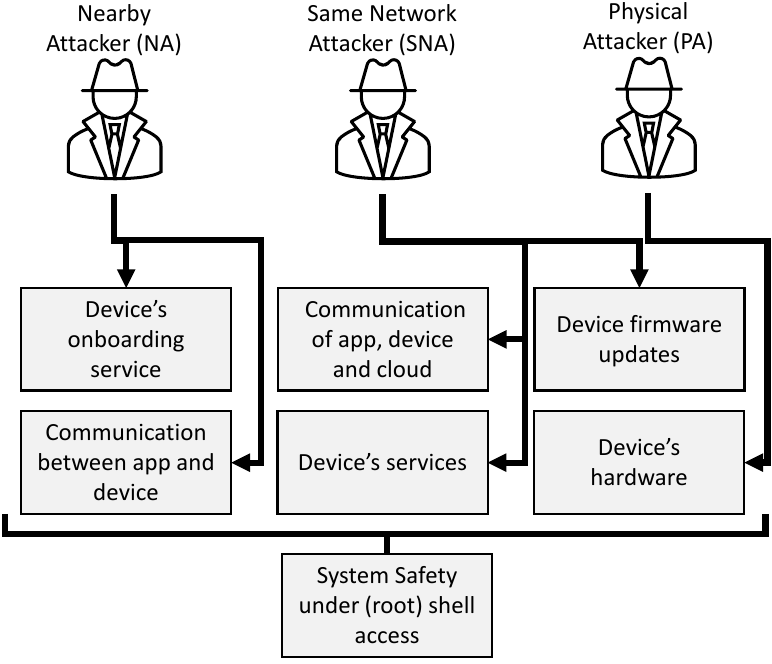}
    \caption{Adversary's relation to attack scenarios in a \smarthome{} setting}
    \label{fig:relation}
    \vspace{-1em}
\end{figure}

\begin{table}[]
\centering
\caption{Attack scenarios for different attacker types}
\begin{scriptsize}
\begin{tabularx}{\linewidth}{X}
\toprule
\spacedlowsmallcaps{(1) Nearby Attacker (NA) -- Communication between the App and the Device}
\\ \midrule
Determine whether the onboarding communication channel uses strong credentials. (\ref{TC-DSP-DefaultSecurity})\newshorttag{Strong Onboarding Channel Credentials}{NA-StrongCredT}
Check whether the data transferred between the app and the device is protected by standardized protocols like TLS/\-SSL. (\ref{TC-STC-SecureTransit})\newshorttag{Onboarding Traffic Protection}{NA-ProtocolsT}
Check whether the data transferred between the app and the device is additionally signed. (\ref{TC-STC-Singed})\newshorttag{Signed Onboarding Traffic }{NA-SignedT}
\\ \midrule
\spacedlowsmallcaps{(2) Nearby Attacker (NA) -- Device's Onboarding Services}
\\ \midrule
Check that default passwords for user accounts are changed and only strong passwords can be set. (\ref{TC-AE-StrongPW})\newshorttag{Strong Device Passwords}{NA-StrongPWT}
Verify that measures against brute-force attacks are applied to secure the device's onboarding services. (\ref{TC-AE-BruteForce})\newshorttag{Brute Force Protection}{NA-BruteForceT}
Determine if the device verifies that it only communicates with the owner's smartphone app.  (\ref{TC-STC-Verified})\newshorttag{Verified Communication Partner}{NA-VerifiedT}
Scan the device for unnecessarily open ports and services that could lead to security issues with \nmap{}. (\ref{TC-STC-Ports})\newshorttag{Unnecessarily Open Ports}{NA-PortsT}
Check the operability of the device and the onboarding services under \ac{DoS} attacks (Ping of Death and \ac{ICMP} Flood Attack). (\ref{TC-STC-LimitBandwidth})\newshorttag{\ac{DoS} Attack Resilience}{NA-LimitBandwidthT}
Analyze onboarding web interfaces and services for common web vulnerabilities. (\ref{TC-SIS-WebVulnerabilities})\newshorttag{Web Vulnerabilities}{NA-WebVulnerabilitiesT}
Check whether manipulated inputs from the app are properly sanitized. (\ref{TC-SIOH-Filter})\newshorttag{Input Sanitization}{NA-FilterT}
Verify that the onboarding network and services are correctly terminated after the onboarding process. (\ref{TC-SIS-NetworkInterfaces})\newshorttag{Onboarding Network Termination}{NA-NetworkInterfaceT}
\\ \midrule
\spacedlowsmallcaps{(3) Same Network Attacker (SNA) -- Communication of App, Device and Cloud}
\\ \midrule
Check whether the data transferred between the app and the cloud is protected by standardized protocols like TLS/SSL and certificate authority verification or \sslpinning{} is applied. (\ref{TC-STC-SecureTransit})\newshorttag{App Traffic Protection}{SNA-SecureTransitAppT}
Check whether the data transferred between the app and the cloud is additionally signed.    (\ref{TC-STC-Singed})\newshorttag{Signed App Traffic}{SNA-SignedAppT}
Check whether the data transferred between the device and the cloud is protected by standardized protocols like TLS/SSL and certificate authority verification or \sslpinning{} is applied. (\ref{TC-STC-SecureTransit})\newshorttag{Device Traffic Protection}{SNA-SecureTransitDeviceT}
Check whether the data transferred between the device and the cloud is additionally signed. (\ref{TC-STC-Singed})\newshorttag{Signed Device Traffic}{SNA-SignedDeviceT}
Determine if 2FA is available for the cloud account or device access. (\ref{TC-AE-2FA})\newshorttag{2FA Authentication}{SNA-2FAT}
\\ \midrule
\spacedlowsmallcaps{(4) Same Network Attacker (SNA) -- Device's Services}
\\ \midrule
Verify that measures against brute-force attacks have been applied to secure the device's services. (\ref{TC-AE-BruteForce})\newshorttag{Brute Force Protection}{SNA-BruteForceT}
Determine if the device verifies that it communicates only with the cloud or smartphone app. (\ref{TC-STC-Verified})\newshorttag{Verified Communication Partner}{SNA-VerifiedT}
Scan the device for unnecessarily open ports and services that could lead to security issues with \nmap{}. (\ref{TC-STC-Ports})\newshorttag{Unnecessarily Open Ports}{SNA-PortsT}
Determine whether the services or endpoints use strong credentials. (\ref{TC-DSP-DefaultSecurity})\newshorttag{Strong Administrative Credentials}{SNA-DefaultSecurityT}
Check the operability of the services under \ac{DoS} attacks (Ping of Death and \ac{ICMP} Flood Attack). (\ref{TC-STC-LimitBandwidth})\newshorttag{\ac{DoS} Attack Resilience}{SNA-LimitBandwidthT}
Analyze web interfaces and services for common web vulnerabilities. (\ref{TC-SIS-WebVulnerabilities})\newshorttag{Web Vulnerabilities}{SNA-WebVulernabilitiesT}
Determine the operability of the device when shutting down the internet connection, e.g., by sink-holing all traffic to external services. (\ref{TC-SSR-RelyNetwork})\newshorttag{Cloudless Service}{SNA-RelyNetworkT}
\\ \midrule
\spacedlowsmallcaps{(5) Same Network Attacker (SNA) -- Device Firmware Updates}
\\ \midrule
Determine whether the device uses an automatic firmware update mechanism. (\ref{TC-FU-UpdateFU})\newshorttag{Automatic Firmware Updates}{SNA-UpdateFUT}
Check whether the firmware update is transmitted via a secure channel. Additionally, the update should be encrypted, signed, and the integrity verified before applying it.  (\ref{TC-FU-SecureFU}, \ref{TC-FU-Encrypt}, \ref{TC-FU-Verify})\newshorttag{Secure Firmware Updates}{SNA-SecureFUT}
Determine whether there are hard-coded passwords or credentials in the firmware update file. (\ref{TC-FU-Passwords})\newshorttag{Firmware Update without hardcoded default Credentials}{SNA-PasswordsT}
Check whether a rollback to a previous version is possible, e.g., by serving or injecting an older firmware update into the cloud communication. (\ref{TC-FU-Rollback})\newshorttag{Firmware Downgrade Prevention}{SNA-RollbackT}
\\ \midrule
\spacedlowsmallcaps{(6) Physical Attacker (PA) -- Device's Hardware}
\\ \midrule
Is the device's packaging tamper-evident? (\ref{TC-HW-Seal})\newshorttag{Tamper-Evident Packaging}{PA-SealT}
Check whether special measures against physical tampering have been applied, like \ac{PCB} coating based on epoxy. (\ref{TC-HW-TamperCase})\newshorttag{Physical Tamper Protection}{PA-TamperCaseT}
Determine whether the device uses special security chips. (\ref{TC-HW-SecureChip})\newshorttag{Security Chips}{PA-SecureChipT}
Analyze all ports and  test points for possible communication interfaces, e.g., \ac{UART} interfaces. (\ref{TC-HW-Ports})\newshorttag{Physical Communication Interfaces}{PA-PortsT}
Check whether interfaces are accessible without any default credentials. (\ref{TC-DSP-DefaultSecurity})\newshorttag{Secure Physical Communication Interface Credentials}{PA-DefaultSecurityT}
Check whether credentials like the \wifi{} password are properly deleted after a factory reset by reading the flash memory after a reset. (\ref{TC-DPC-DataFactoryReset})\newshorttag{Factory Reset deletes private data}{PA-DataFactoryResetT}
Check whether the file system is encrypted by reading out the flash memory. (\ref{TC-FU-EncryptedFS})\newshorttag{Encrypted File System}{PA-EncryptedFST}
\\ \midrule
\spacedlowsmallcaps{(7) System Safety under (Root) Shell Access}
\\ \midrule
Determine whether the device implements secure boot functionalities based on security chips. (\ref{TC-HW-SecureChip})\newshorttag{Secure Boot Functionality}{SA-SecureChipT}
Check whether credentials are securely stored on the device. (\ref{TC-AE-IndustryHash})\newshorttag{Proper Credential Protection}{SA-IndustryHashT}
Determine what kind of privileges services running on the device have. (\ref{TC-AO-LeastPriviledge})\newshorttag{Service Privileges}{SA-LeastPriviledgeT}
Determine if users can write to the file system.  (\ref{TC-AO-RootFS})\newshorttag{User Writable File System}{SA-RootFST}
Is the device using a firewall? (\ref{TC-SIS-Firewall})\newshorttag{Firewall Protection}{SA-FirewallT}
Search for a logging system on the device. If there is one, check that no sensitive data is logged. (\ref{TC-LOG-Filter})\newshorttag{No logging of sensitive data}{SA-FilterT}
Determine if there are any user alerts about successful security breaches. (\ref{TC-MAT-Inform})\newshorttag{Security Breach Alerts}{SA-InformT}
\\ \bottomrule
\end{tabularx}
\end{scriptsize}
\label{tab:scenario}
\end{table}
			The Ping of Death \ac{DoS} attack is executed by a custom Python script that sends fragmented IP packets having a larger size than 65,535 bytes.
			The second \ac{DoS} attack, the \ac{ICMP} Flood attack, is tested by sending 65,535-byte \ac{ICMP} pings from 100 terminals for one minute.
			Both \ac{DoS} attacks succeed (and the test case fails) if the device is not reachable anymore, e.g., by the smartphone app during or after the attack.
			The analysis of the smartphone app for general information retrieval is performed with jadx\footnote{\url{https://github.com/skylot/jadx}} and apktool\footnote{\url{https://ibotpeaches.github.io/Apktool/}}.

			Ideally, the onboarding network is shut down after the entire configuration process and it is not possible to perform command injection attacks via the smartphone app or manipulated requests.

		\subsection{\acl{SNA} Test Cases}\label{sec:TestSuite:SameNetwork}

			The attacker on the same network can target all services of the \smarthome{} Device that are accessible to the internal network after the initial setup.
			The attacker can also intercept and manipulate all traffic between the device and the \ac{IoT} cloud.
			Furthermore, traffic originating from the smartphone app can also be analyzed by performing a \ac{MitM} attack, e.g., by manipulating the local \ac{DNS} entries (\ac{DNS} Spoofing) or \ac{ARP} Spoofing.
			The list of test cases for this attacker type contains cases from the nearby attacker and in addition test cases from the \nameref{sec:TestCaseSelection:SystemSafety} and \nameref{sec:TestCaseSelection:Firmware} categories.

			The test cases defined in \autoref{tab:scenario} (3) focus on the cloud communication of the device and the smartphone app.
			The traffic from the smartphone app and the device to the cloud is again analyzed with Wireshark, Burp Suite, sslyze and in addition mitmproxy\footnote{\url{https://mitmproxy.org/}} to automatically generate SSL certificates for the accessed domain.
			To investigate the security measure for certificate authority checks or \sslpinning{} of the app, Frida\footnote{\url{https://frida.re/docs/home/}} \& Objection\footnote{\url{https://github.com/sensepost/objection}} is used in conjunction with jadx and apktool.

			The test cases in \autoref{tab:scenario} (4) investigate the locally exposed services and ports.
			Here again, brute-force measures are successful if only a few requests are possible or are completely blocked.
			If the device does not support local streaming, the device should only communicate with the cloud and not have any unnecessarily open ports.
			For administrative endpoints, the same recommendations as for strong passwords from Grassi et al.~\cite{Grassi2017} apply.
			\ac{DoS} attacks are executed as already described in Sect.~\ref{sec:TestSuite:Nearby} and \ac{DNS} Spoofing is done by directly setting \ac{DNS} records on the local \ac{DNS} server.

			Last but not least, the firmware update process is analyzed using the test cases in \autoref{tab:scenario} (5).
			At a minimum, it must be possible to enable automatic firmware updates and the firmware update must be transmitted over a secure channel that prevents  manipulation attacks.
			Additionally, the firmware update must not contain hard-coded credentials and a rollback to a previous firmware version should be prevented.
			This is important to mitigate a rollback to versions with known security vulnerabilities.

		\subsection{\acl{PA} Test Cases}\label{sec:TestSuite:Physical}

			The \acl{PA} targets physical ports, debug interfaces, hardware and the encasing of the device.
			The capabilities of this type of attacker include disassembling the device and (de-)soldering individual hardware components.
			Therefore, test cases from the \nameref{sec:TestCaseSelection:Hardware}, \nameref{sec:TestCaseSelection:DefaultSecurity}, \nameref{sec:TestCaseSelection:Protection} and \nameref{sec:TestCaseSelection:Firmware} sections are relevant.

			\autoref{tab:scenario} (6) shows the attack scenario for the physical hardware of the device.
			At first, the tamper protection is investigated and it is checked whether tampering with the packaging and the device, e.g., during the shipment process can be detected.
			Then all important components and chips on the \ac{PCB} are identified and checked for special security functions.
			Last but not least, test points, debugging interfaces, and connectors are analyzed.
			For this analysis, a multimeter, oscilloscope, and a CH341A USB Programmer are used to access \ac{UART} interfaces and flash memory.
			The extraction of file systems from flash memory is accomplished with flashrom\footnote{\url{https://www.flashrom.org/Flashrom}} or the UsbAsp-flash\footnote{\url{https://github.com/nofeletru/UsbAsp-flash}} software in combination with the CH341A USB Programmer.
			Binary files obtained from firmware update files or dumped \ac{SPI} chips are analyzed with binwalk\footnote{\url{https://github.com/ReFirmLabs/binwalk}}, IDA Pro\footnote{\url{https://hex-rays.com/IDA-pro}} and Ghidra\footnote{\url{https://ghidra-sre.org/}}.

		\subsection{\acl{SA} Test Cases}\label{sec:TestSuite:Shell}

			Some test cases defined in \autoref{sec:app3} require at least shell access to the \smarthome{} Device.
			Therefore, the following test cases in \autoref{tab:scenario} (7) are only executed at least one of the preceding test cases and were able to gain (root) shell access to the device.
			This can happen, e.g., by an improperly protected \ac{UART} interface, \telnet{} services offering a shell login or command injection vulnerabilities.

			This attack scenario focuses on various aspects of the operating system and whether measures are applied to increase the system security level.
			Such measures include the use of a secure boot functionality based on security chips, secure storing of credentials, and the use of a firewall.
			In addition, services running on the device and user accounts should have only the least privileges needed and should not run with root capabilities.
			The system is also searched for a logging system and whether that system stores sensitive data in an unprotected manner.
			In the ideal case, there should be a notification on the users' smartphones informing them of a possible security breach after someone gains shell access to the device.
			
			All defined test cases and attackers form the basis for our security analysis of nine real-world \smarthome{} Devices presented in the following section.
		
\section{Evaluation}
\label{sect:evaluation}

We evaluate the \aclp{BP} using a selection of different \wifi{} \smarthome{} cameras, a \wifi{} \smarthome{} Vacuum Cleaner, and a \wifi{} Smoke \& CO Alarm Detector.
	The \smarthome{} cameras are manufactured by well-known vendors and can be divided into two groups.
	The first group consists of high-end devices with a relatively high price ($\geq$~\$99.00): the \emph{\arlocam}, \emph{\boschcam}, \emph{\nestcam} and \emph{\ringcam}.
	Devices in the second group are significantly cheaper (\$30.00 -- \$55.00) and can be considered mid-range \smarthome{} cameras.
	This group comprises \emph{\blinkcam}, \emph{\dlinkcam} and \emph{\tpcam}.
	The latter two are the only cameras that can be used independently of their vendor's cloud service.
	The other devices include the \emph{\nestprotect{}} and \emph{\tesvor{}}.
	Both devices do not necessarily need a cloud connection to function, but the vendor's cloud is required for access to the device via the smartphone app or for receiving push notifications.

For each of the above devices the gathered test suites are executed, starting with the Information Gathering Phase outlined in \autoref{sec:information_gathering}, after which
the test cases of the Nearby Attacker and the Same Network Attacker are executed.
Finally, the Physical Attacker Test Cases are applied. In case any of the attacks were successful in gaining shell access to the device, the test is concluded by executing the Shell Access Test Cases.
For all discovered vulnerabilities we calculate corresponding  \ac{CVSS} base scores, which reflect the severity of each vulnerability.

\subsection{Analysis Findings}
An overview of the detailed analysis results is included in \autoref{tab:AnalysisSummary1} and \autoref{tab:AnalysisSummary2} in \autoref{sec:app2} for each attacker type and device.
The main findings are discussed in the following.

Our evaluation shows that the main issues during the onboarding process are related to communication channels that are not properly protected, e.g., with appropriate credentials and encryption standards.
Additionally, we found that for none of the tested devices, the onboarding traffic is  signed and that the communication partners are not properly verified, allowing a potential \acl{NA} to communicate with the onboarding services.
		It also indicates that especially the traffic originating from the app could leverage signing techniques more often and devices offering a service should verify communication partners, e.g., based on certificates.
		Furthermore, all devices (except the \boschcam{}) that provide services on the network lack brute-force protection and allow someone on the same network to attack services with default or only weak credentials.

		Concerning physical attacks, every device has at least one weak point.
		The majority of the devices lack tamper-evident packaging, which would allow them to identify manipulations.
		Moreover, all \acp{PCB} and cases, with the exception of the \ringcam{} do not have tamper protection and do not hinder a physical attacker from interacting with the physical interfaces.
		This is especially an issue for devices that have an active debugging interface on their \ac{PCB} and because all analyzed flash memory chips (6 out of 9) did not use file system encryption.

		Detailed information about the operating system could be retrieved for the Arlo, D-Link, and TP-Link cameras.
		These three cameras do not follow the recommendations for securing the operating system and, for example, do not properly protect credentials, run services with root privileges, and do not use a properly configured firewall to reduce the attack surface.
		In addition, the \dlinkcam{} and \tpcam{} do not have a mechanism to identify security breaches and notify the owner about successful attacks.

		The results of the security analysis can be summarized as follows.
		Four\footnote{\blinkcam{}, \dlinkcam{}, \tesvor{}, \tpcam{}} of the nine devices ($\approx$44\%) have at least one security vulnerability with a medium ($\geq$ 4.0) or high ($\geq$ 7.0) rated \ac{CVSS} base score.
		All of these vulnerabilities directly threaten the user's privacy and the integrity of the device by, for example, leaking \wifi{} credentials, enabling \ac{MitM} attacks, or relying on weak default passwords.
		The \dlinkcam{} and the \tpcam{}, have even three or more different vulnerabilities rated medium, high, or critical ($\geq$ 9.0).
		In particular, the critical vulnerability in the \tpcam{} allows an attacker who is on the same network or within range of the onboarding \wifi{} network to easily gain root access.
		In contrast to that five devices\footnote{\arlocam{}, \boschcam{}, \nestcam{}, \nestprotect{}, \ringcam{}} ($\approx$56\%) have no direct security issue or only low-rated vulnerabilities and can therefore be considered secure.
		However, as the evaluation of the \acl{BP} shows, the security of all devices suffers from not following \ac{IoT} security \aclp{BP}.
		The main issues for all devices are missing file system encryption, proper communication partner verification, tamper-protection, and protection of the onboarding process with industry-standard encryption.
		As a result, it can be stated that following the identified \ac{IoT} security \aclp{BP} would have prevented most or even all vulnerabilities and would have increased the robustness against attacks.	

	\subsection{Analysis of Best Practices and Vulnerabilities}

	In order to quantify the security impact of specific \aclp{BP}, we identify such \aclp{BP}, for which non-adherence in tested devices frequently leads to security vulnerabilities, and calculate their associated average \ac{CVSS} base score.
    The results are shown in \autoref{tab:BPAnalysis}.
	In addition, the test cases that contributed to the discovery of the vulnerabilities are listed for each vulnerability.
	The link between the vulnerability and the best practice can be established from the results of the detailed security analysis results in \autoref{sec:app2}.

\begin{figure}
    \centering
    \includegraphics[width=0.95\linewidth, trim={11.9cm 4.7cm 11.9cm 6cm},clip]{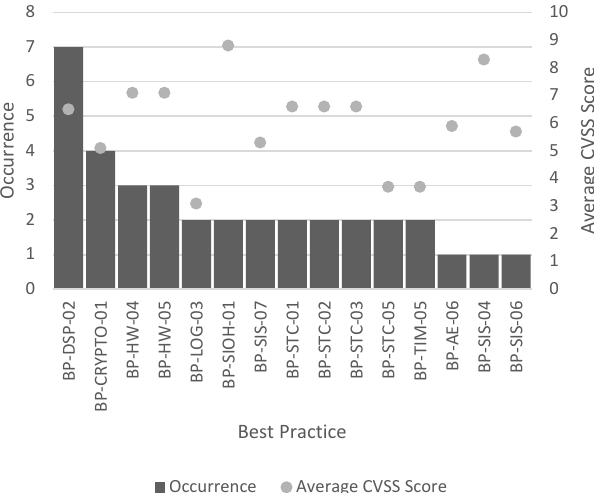}
    \caption{Occurrences and average \ac{CVSS} Scores of \aclp{BP} leading to Vulnerabilities}
    \label{fig:AnalysisBPCausingVuln}
    \vspace{-1em}
\end{figure}

	The results in \autoref{tab:BPAnalysis} form the basis for a quantitative analysis of \aclp{BP} that led to the vulnerabilities shown in \autoref{fig:AnalysisBPCausingVuln} (shown in more detail in \autoref{tab:AnalysisBPCausingVuln} in \autoref{sec:app2}).
	It is sorted by the occurrence of each \acl{BP} in \autoref{tab:BPAnalysis} and shows that \ref{BP-Default-PWCrack} is the \acl{BP} which is connected to most of the security vulnerabilities (7 out of 16).
	\ref{BP-Default-PWCrack} encourages the vendor to use individual strong default passwords that are hard to guess.
	By looking at the vulnerabilities in \autoref{tab:BPAnalysis} it can be seen that this \acl{BP} is connected to different attack vectors which have an average \ac{CVSS} base score of 6.5.
	These vectors are related to onboarding networks, local home networks, and physical vulnerabilities and are often caused by either not using a default password at all or only using weak, easy-to-guess passwords.
	In \autoref{tab:BPAnalysis} it can be observed that hardware security vulnerabilities (\ref{BP-HW-RemoveDebug}, \ref{BP-HW-AdminInter}) are always present in combination with \ref{BP-Default-PWCrack} which occurred three times in total (see \autoref{fig:AnalysisBPCausingVuln}).
	The reason for this is that the vendors did not disable the debug interfaces in the two affected devices and did not protect them physically.
	In combination with a weak or no (default) root password (\ref{BP-Default-PWCrack}), this chain leads to root access with an average \ac{CVSS} base score of 7.1.
 
		The \acl{BP} \ref{BP-CRYPTO-Proto} occurs also frequently. It states that the strongest cryptographic protocol should be used whenever possible when exchanging data and connecting to other services to protect confidentiality, authenticity, and integrity.
		This \acl{BP} is associated with four vulnerabilities where insufficient authentication allows an adversary to communicate and tamper with the device during the configuration phase, or, where inadequate encryption protocols cause a breach of confidentially and integrity.
		The average \ac{CVSS} base score for this \acl{BP} is 5.1, others occur only once or twice.

		By taking a closer look at the underlined high ($\geq$ 7.0) average \ac{CVSS} base scores in \autoref{fig:AnalysisBPCausingVuln} and in addition \autoref{tab:BPAnalysis}, it can be concluded that these four \aclp{BP} are always connected to vulnerabilities that allow root access.
		Especially \acl{BP} \ref{BP-SIOH-Filter} that focuses on secure input handling and has the highest average \ac{CVSS} base score is associated with two root command injection vulnerabilities in the \dlinkcam{} and the \tpcam{}.
		This suggests that vendors should specifically focus on proper input validation (\ref{BP-SIOH-Filter}), disabling unnecessary services and interfaces (\ref{BP-SINS-NetworkInterfaces}), and removing or securing physical administrative interfaces (\ref{BP-HW-RemoveDebug}, \ref{BP-HW-AdminInter}).
		On the other hand, logging of credentials (\ref{BP-LOG-Credentials}) and not verifying the communication partner (\ref{BP-STC-Verified}, \ref{BP-TIM-Relationships}) only result in low average \ac{CVSS} base scores ($<$ 4.0).
		However, there are several medium-rated \aclp{BP} that should be followed because they can lead in individual cases to serious confidentiality breaches  such as the missing SSL certificate validation in the \blinkcam{}.

		\begin{table}[htbp]
			\myfloatalign
			\caption{Vulnerabilities and their underlying \acl{BP}}
			\label{tab:BPAnalysis}
			\begin{scriptsize}
				\begin{tabularx}{\linewidth}{lXXX}
					\toprule
					\tableheadline{Device}                                       & \tableheadline{Vulnerability}                                                                                                                                                                                 & \tableheadline{Test Case}                                                          & \tableheadline{Best Practice}                                                                                               \\
					\midrule
					Arlo\hyperlink{tab:bpresults:arlo}{\textsuperscript{1}}      & Onboarding takeover with different user possible (\ac{CVSS} \href{https://nvd.nist.gov/vuln-metrics/cvss/v3-calculator?vector=AV:A/AC:H/PR:N/UI:R/S:U/C:L/I:N/A:L\&version=3.1}{3.7})                         & \ref{TC-STC-Verified}                                                              & \ref{BP-TIM-Relationships}, \ref{BP-CRYPTO-Proto}, \ref{BP-STC-Verified}                                                    \\ \hline
					Arlo\hyperlink{tab:bpresults:arlo}{\textsuperscript{1}}      & Logging of sensitive data, e.g., access tokens (\ac{CVSS} \href{https://nvd.nist.gov/vuln-metrics/cvss/v3-calculator?vector=AV:P/AC:H/PR:N/UI:N/S:U/C:L/I:N/A:N\&version=3.1}{2.0})                            & \ref{TC-LOG-Filter}                                                                & \ref{BP-LOG-Credentials}                                                                                                    \\ \hline
					Blink\hyperlink{tab:bpresults:blink}{\textsuperscript{2}}    & \ac{DoS} attacks possible (\ac{CVSS} \href{https://nvd.nist.gov/vuln-metrics/cvss/v3-calculator?vector=AV:A/AC:L/PR:N/UI:R/S:U/C:N/I:N/A:H\&version=3.1}{5.7})                                                & \ref{TC-STC-LimitBandwidth}                                                        & \ref{BP-SINS-DDoS}                                                                                                          \\ \hline
					Blink\hyperlink{tab:bpresults:blink}{\textsuperscript{2}}    & Missing SSL certificate verification leads to \ac{MitM} attack vectors (\ac{CVSS}~\href{https://nvd.nist.gov/vuln-metrics/cvss/v3-calculator?vector=AV:N/AC:H/PR:N/UI:N/S:U/C:H/I:N/A:H\&version=3.1}{7.4})   & \ref{TC-STC-SecureTransit}                                                         & \ref{BP-CRYPTO-Proto}, \ref{BP-STC-Transit}, \ref{BP-STC-SecurityProto}, \ref{BP-STC-Credentials}                           \\ \hline
					D-Link\hyperlink{tab:bpresults:dlink}{\textsuperscript{3}}   & Command injection vulnerabilities in \bluetooth{} services (\ac{CVSS} \href{https://nvd.nist.gov/vuln-metrics/cvss/v3-calculator?vector=AV:A/AC:H/PR:N/UI:N/S:C/C:H/I:H/A:H\&version=3.1}{8.3})               & \ref{TC-DSP-DefaultSecurity}, \ref{TC-SIS-NetworkInterfaces}, \ref{TC-SIOH-Filter} & \ref{BP-Default-PWCrack}, \ref{BP-SINS-NetworkInterfaces}, \ref{BP-SIOH-Filter}                                             \\ \hline
					D-Link\hyperlink{tab:bpresults:dlink}{\textsuperscript{3}}   & Weak unchangeable \ac{RTSP} service password	(\ac{CVSS} \href{https://nvd.nist.gov/vuln-metrics/cvss/v3-calculator?vector=AV:A/AC:L/PR:N/UI:N/S:U/C:H/I:N/A:N\&version=3.1}{6.5})                              & \ref{TC-DSP-DefaultSecurity}                                                       & \ref{BP-Default-PWCrack}                                                                                                    \\ \hline
					D-Link\hyperlink{tab:bpresults:dlink}{\textsuperscript{3}}   & Unauthenticated \ac{RTSP} access via \lstinline[basicstyle=\scriptsize]|StreamProxy| binary (\ac{CVSS} \href{https://nvd.nist.gov/vuln-metrics/cvss/v3-calculator?vector=AV:A/AC:H/PR:N/UI:N/S:U/C:H/I:N/A:N\&version=3.1}{5.3}) & \ref{TC-SIS-WebVulnerabilities}                                                    & \ref{BP-SINS-UserSession}                                                                                                   \\ \hline
					D-Link\hyperlink{tab:bpresults:dlink}{\textsuperscript{3}}   & Unauthenticated \ac{UART} interface (\ac{CVSS} \href{https://nvd.nist.gov/vuln-metrics/cvss/v3-calculator?vector=AV:P/AC:H/PR:N/UI:N/S:C/C:H/I:H/A:H\&version=3.1}{7.1})                                      & \ref{TC-HW-Ports}, \ref{TC-DSP-DefaultSecurity}                                    & \ref{BP-HW-RemoveDebug}, \ref{BP-HW-AdminInter}, \ref{BP-Default-PWCrack}                                                   \\ \hline
					D-Link\hyperlink{tab:bpresults:dlink}{\textsuperscript{3}}   & \telnet{} activation via MicroSD Card file (\ac{CVSS} \href{https://nvd.nist.gov/vuln-metrics/cvss/v3-calculator?vector=AV:P/AC:H/PR:N/UI:N/S:C/C:H/I:H/A:H\&version=3.1}{7.1})                               & \ref{TC-HW-Ports}, \ref{TC-DSP-DefaultSecurity}                                    & \ref{BP-HW-RemoveDebug}, \ref{BP-HW-AdminInter}, \ref{BP-Default-PWCrack}                                                   \\ \hline
					D-Link\hyperlink{tab:bpresults:dlink}{\textsuperscript{3}}   & Logging of sensitive data, e.g., on \ac{UART} interface (\ac{CVSS} \href{https://nvd.nist.gov/vuln-metrics/cvss/v3-calculator?vector=AV:P/AC:H/PR:N/UI:N/S:U/C:H/I:N/A:N\&version=3.1}{4.2})                   & \ref{TC-LOG-Filter}                                                                & \ref{BP-LOG-Credentials}                                                                                                    \\ \hline
					Ring\hyperlink{tab:bpresults:ring}{\textsuperscript{4}}      & Lack of onboarding service authentication measures (\ac{CVSS} \href{https://nvd.nist.gov/vuln-metrics/cvss/v3-calculator?vector=AV:A/AC:H/PR:N/UI:R/S:U/C:L/I:N/A:L\&version=3.1}{3.7})                       & \ref{TC-DSP-DefaultSecurity}, \ref{TC-STC-Verified}                                & \ref{BP-Default-PWCrack}, \ref{BP-TIM-Relationships}, \ref{BP-CRYPTO-Proto}, \ref{BP-STC-Verified}                          \\ \hline
					Tesvor\hyperlink{tab:bpresults:tesvor}{\textsuperscript{5}}  & Weak password recovery process with 4-digit PIN	(\ac{CVSS} \href{https://nvd.nist.gov/vuln-metrics/cvss/v3-calculator?vector=AV:N/AC:H/PR:N/UI:N/S:U/C:H/I:N/A:N\&version=3.1}{5.9})                           & ---                                                                                & \ref{BP-AUTHEN-PWReset}                                                                                                     \\ \hline
					Tesvor\hyperlink{tab:bpresults:tesvor}{\textsuperscript{5}}  & Leakage of \wifi{} credentials during onboarding	(\ac{CVSS} \href{https://nvd.nist.gov/vuln-metrics/cvss/v3-calculator?vector=AV:A/AC:L/PR:N/UI:R/S:U/C:H/I:N/A:N\&version=3.1}{5.7})                          & \ref{TC-DSP-DefaultSecurity}, \ref{TC-STC-SecureTransit}                           & \ref{BP-Default-PWCrack}, \ref{BP-CRYPTO-Proto}, \ref{BP-STC-Transit}, \ref{BP-STC-SecurityProto}, \ref{BP-STC-Credentials} \\ \hline
					TP-Link\hyperlink{tab:bpresults:tplink}{\textsuperscript{6}} & \ac{RTSP} authentication bypass in \lstinline[basicstyle=\scriptsize]|cet| binary (\ac{CVSS} \href{https://nvd.nist.gov/vuln-metrics/cvss/v3-calculator?vector=AV:A/AC:H/PR:N/UI:N/S:U/C:H/I:N/A:N\&version=3.1}{5.3})   & \ref{TC-SIS-WebVulnerabilities}                                                    & \ref{BP-SINS-UserSession}                                                                                                   \\ \hline
					TP-Link\hyperlink{tab:bpresults:tplink}{\textsuperscript{6}} & Command injection vulnerability in \lstinline[basicstyle=\scriptsize]|uhttpd| binary	(\ac{CVSS} \href{https://nvd.nist.gov/vuln-metrics/cvss/v3-calculator?vector=AV:A/AC:L/PR:N/UI:N/S:C/C:H/I:H/A:H\&version=3.1}{9.3})    & \ref{TC-SIOH-Filter}                                                               & \ref{BP-SIOH-Filter}                                                                                                        \\ \hline
					TP-Link\hyperlink{tab:bpresults:tplink}{\textsuperscript{6}} & Improperly protected \ac{UART} interface with root shell (\ac{CVSS} \href{https://nvd.nist.gov/vuln-metrics/cvss/v3-calculator?vector=AV:P/AC:H/PR:N/UI:N/S:C/C:H/I:H/A:H\&version=3.1}{7.1})                 & \ref{TC-HW-Ports}, \ref{TC-DSP-DefaultSecurity}                                    & \ref{BP-HW-RemoveDebug}, \ref{BP-HW-AdminInter}, \ref{BP-Default-PWCrack}                                                   \\
					\bottomrule
				\end{tabularx}
				\\
				\vspace*{\baselineskip}
				\hypertarget{tab:bpresults:arlo}{\textsuperscript{1}\arlocam{}} \quad \hypertarget{tab:bpresults:blink}{\textsuperscript{2}\blinkcam{}} \quad \hypertarget{tab:bpresults:dlink}{\textsuperscript{3}\dlinkcam{}} \quad \hypertarget{tab:bpresults:ring}{\textsuperscript{4}\ringcam{}} \quad \hypertarget{tab:bpresults:tesvor}{\textsuperscript{5}\tesvor{}} \\
				\hypertarget{tab:bpresults:tplink}{\textsuperscript{6}\tpcam{}}
			\end{scriptsize}
   \vspace{-2.5em}
		\end{table}
\vspace{-1.2em}

\subsection{Analysis Outcomes}

\begin{figure*}[ht]
    \centering
    \includegraphics[width=0.95\textwidth, trim={3.0cm 3.4cm 4.1cm 0.3cm},clip]{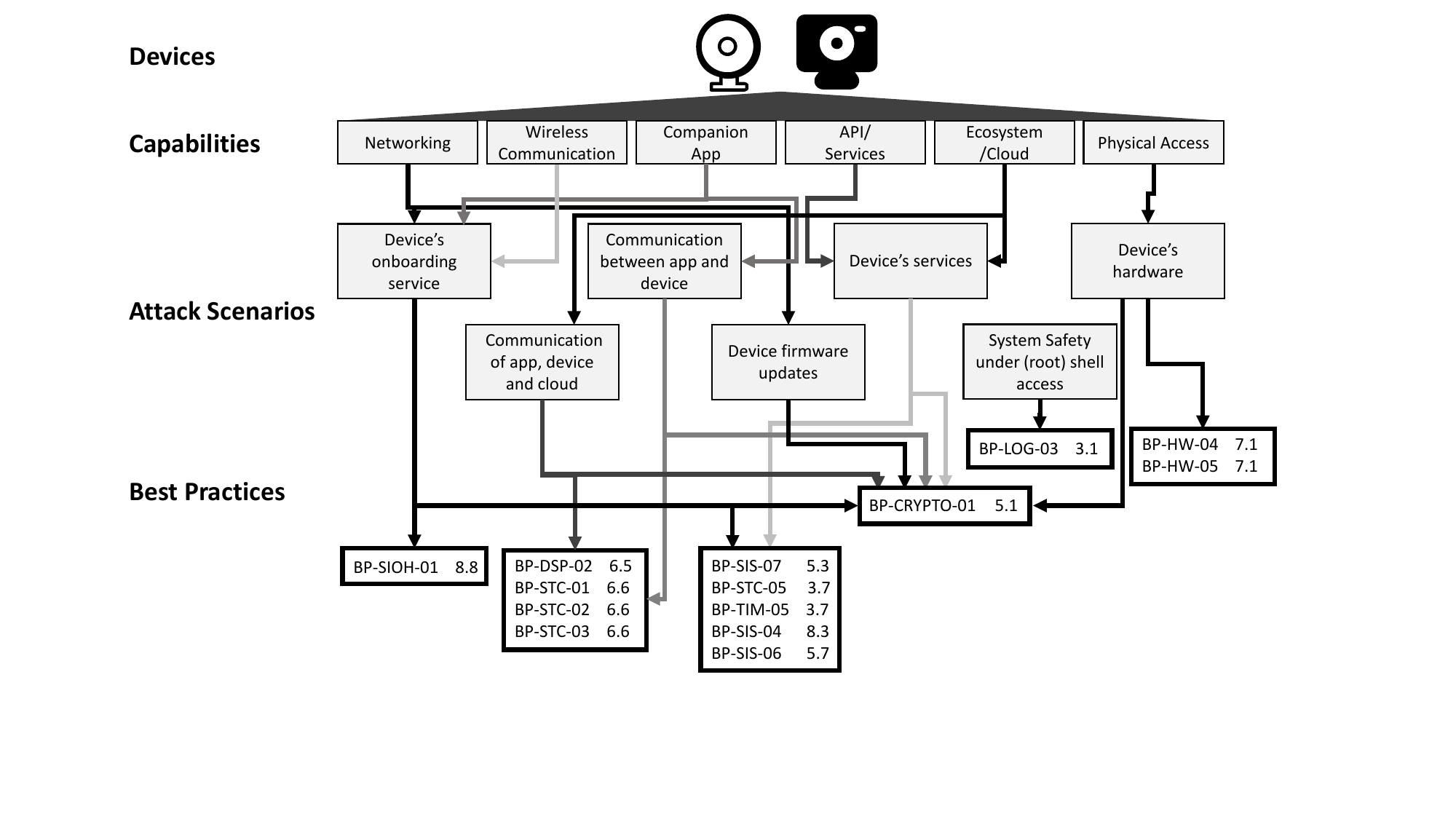}
    \caption{Decision graph for prioritization of \aclp{BP} in relation to different \smarthome{} device capabilities and relevant attack scenarios. \aclp{BP} with their average CVSS scores ordered by occurrence frequency of associated vulnerabilities.}
    \label{fig:decision}
\end{figure*}

Ideally, manufacturers of IoT products should consider \emph{all} \aclp{BP} presented in order to achieve a maximum level of avoidance of security harm associated with their products. In reality, however, due to limited resources and time, vendors often need to make compromises with regard to which \aclp{BP} to consider in their implementations. In order to aid vendors to decide which best practices to prioritize in the implementation of  IoT devices, we give precedence to the most effective best practices in terms of potential reduction of harm and combine our evaluation results in the form of a decision graph shown in \autoref{fig:decision}. It is based on the insights of \autoref{sec:experiment} (attack scenarios) and the weighted Best Practices evaluated in \autoref{sect:evaluation}. Depending on the different capabilities of a consumer IoT device, namely the presence of network or wireless communication interfaces, availability of a companion app, device-based APIs or services, or, a cloud-based service ecosystem, and potential physical exposure of the device in question, the attacker has access to different attack surfaces related to distinct attack scenarios.
Depending on which attack scenarios are relevant for a specific IoT device based on its features and capabilities, the decision tree shows a mapping of each attack scenario to a listing of \aclp{BP} that are particularly important to follow in each scenario.

The decision graph in \autoref{fig:decision} is a mapping from device capabilities to relevant attack scenarios and preferred best practices to consider. For each scenario, a list of \aclp{BP} is provided, ordered by the occurrence frequency of vulnerabilities related to the \acl{BP} alongside associated average CVSS scores. Note that, as described in \autoref{sec:TestSuite:Shell} and \autoref{fig:relation}, the attack scenario of \enquote{System Safety under (root) shell access} is relevant for any scenarios which may potentially lead to a situation in which (root) shell access is available to the attacker.

Consider, as an example, a typical IoT device with \textit{networking capabilities}. For such devices, at least attack scenarios related to the \textit{device's onboarding service} and \textit{device firmware updates} are relevant and need to be considered. For these attack scenarios \aclp{BP} \ref{BP-SIOH-Filter} (CVSS score 8.8) and \ref{BP-CRYPTO-Proto} (CVSS score 5.1) are particularly important to consider and should be prioritized in the design and testing of the device.

In a similar way, for any consumer IoT devices with particular capabilities, vendors can utilize the decision graph to identify the set of \aclp{BP} to consider in their device implementations.

\section{Related Work}\label{sec:related}
To the best of our knowledge no prior work has proposed a methodology for evaluating the security impact of IoT \aclp{BP} and prioritizing and ranking them based on their effectiveness in protecting against security threats on real-world commodity IoT devices.
Therefore, in the following, we concentrate on providing an excerpt of existing attacks on \ac{IoT} devices as well as related scoring systems and security frameworks.

\paragraph{Attacks on Comoodity IoT Devices}
	Paleari~\cite{Paleari2011} reported multiple vulnerabilities in IP cameras from \textit{TRENDnet}, \textit{Digicom}, and \textit{iPUX}. The affected firmware versions allowed attackers to gain administrative access via an undocumented user and escalate privileges to root access by using a command injection vulnerability. Heffner~\cite{Heffner2013} also presented multiple zero-day vulnerabilities, which led to root access in over 50 consumer and professional network camera devices. Additionally, he presented a trivial solution to freeze or replace the live video stream.

	In the same year Crowley et al.~\cite{Crowley2013} presented multiple root access and backdoor vulnerabilities in network-connected embedded devices. They discovered unauthenticated endpoints, path traversal attacks, firmware update attacks, \ac{SSRF} and \ac{CSRF} vulnerabilities in products like \textit{Belkin's WeMo} \wifi{} switch, home automation control units, smart toilets and a smart 'rabbit' from \textit{Karotz}, which can be used as a \wifi{} camera.
	Calmejane et al.~\cite{Calmejane2013} discovered a privilege escalation vulnerability of \textit{Foscam} IP camera devices, which allowed Shekyan et al.~\cite{Shekyan2013} to analyze the firmware version of the affected IP cameras and discover a new \ac{CSRF} attack, to create arbitrary accounts which then could be used to access the camera.

	Tekeoglu et al.~\cite{Tekeoglu2015} revealed multiple security issues ranging from default administrative credentials and open \telnet{} ports to unencrypted video traffic from the device to the vendor's cloud of Belkin \textit{NetCams}. Via these security issues, it was possible to gain root access through the \telnet{} service with the default credentials.

	Favaretto et al.~\cite{Favaretto2019} showed that the \textit{Sricam SP009} IP camera does not encrypt the communication to the vendor's cloud servers, and also, the process of adding a new device was only protected by the enumerable device ID and the (default) password of the device. This can lead to even unskilled attackers gaining full administrative access to other devices via the app.

	Abdalla et al.~\cite{Abdalla2020} discovered multiple security flaws in a generic \textit{Onvif YY HD} branded camera, such as default credentials and guessable device identifiers combined with the unencrypted transmission of the \wifi{} password and weakly hashed login passwords.

	Bitdefender~\cite{Bitdefender2019}, a \smarthome{} company specialized in security and safety systems, discovered that the \textit{Ring Video Doorbell Pro} leaked \wifi{} credentials during the setup process due to the usage of an open network for the setup process.

	Wardle et al.~\cite{Wardle2014} discovered a bootloader attack that could be executed via an exposed \ac{UART} header on the \acf{PCB} of the cloud-based \wifi{} video camera \textit{Dropcam}, which allowed root access on the device which led to the discovery of a vulnerable openssl version (client-side heartbleed attack) and remote command execution through a vulnerable busybox version.
	They also found a reset button triggering a firmware download mode that allowed flashing a new firmware via the vendor's \textit{DirectUSB} tool.
	The corresponding Dropcam iOS app did in addition not use \sslpinning{} and was thereby vulnerable to \ac{MitM} attacks.

	The smart \textit{Nest Thermostat} also lacked hardware security as shown by Hernandez et al.~\cite{Hernandez2014}, by using a USB connection in combination with a device reset, it was possible to send a custom \textit{x-loader} bootloader which bypassed the original \textit{x-loader} bootloader and boot process.
	This allowed an attacker to gain full access to the device's firmware.

    Michele et al.~\cite{Michele2014} discovered that multiple \textit{Smart TV generations} from Samsung use an outdated FFmpeg version with multiple security vulnerabilities.
	Exploiting one of these via distribution of an exploit video, which is then played via a USB drive on the TV, results in full access to the Smart TVs operating system and integrated camera and microphone.

	Even with perfect physical and network security, it is possible to determine the state of \smarthome{} Devices by analyzing the egress and ingress traffic of a \smarthome{} as Copos et al.~\cite{Copos2016} and Apthorpe et al.~\cite{Apthorpe2017} have shown.
	Given the metadata (destination IP) used in an IP packet, it is possible to identify individual traffic flows and correlate \smarthome{} Device states with traffic patterns, identify the devices in use, detect a change in the state of a device, and classify the devices state.
	Acar et al.~\cite{Acar2020} method even work with WPA encrypted \wifi{}, Bluetooth, and \zigbee{} traffic and can achieve an accuracy of over 90\% in determining the state of \smarthome{} Devices.

    In contrast to related work, our proposed overall methodology consists of multiple attacker types executing attacks which cover the entire attack surface of the analyzed devices. As such, our goal is not to either just gain root access to the device or to retrieve information, as was the goal in some previous related works.
    Moreover, every service of each device is systematically evaluated and weaknesses are indicated without regard to exploitability or severity.
    However, this work does not investigate \ac{IoT} (state) identification or cloud security. However, results from Wardle et al.~\cite{Wardle2014} and Klasmark et al.~\cite{Klasmark2020} about the \textit{Dropcam} and \textit{Nest Indoor Camera} are used as a basis to analyze the camera again.
 
\paragraph{Risk Assessment}
The most utilized Scoring System for assessing the exploitability of (software) vulnerabilities is CVSS~\cite{FIRST}. It returns a base score between 0 and 10 for each vulnerability but fails in describing how secure a device is in its entirety. In addition to CVSS, Bonilla et al.~\cite{bonilla2017metric} compare different approaches and their applicability namely, Mean Time-to-Compromise~\cite{leversage2008estimating}, Vulnerability Exposure~\cite{mcqueen2006time}, VEA-bility~\cite{tupper2008vea}, Lai and Hsia’s Model~\cite{lai2007using}, as well as proposing their own metric.

While such metrics may be helpful in judging if a finished product could have vulnerabilities or how severe these known vulnerabilities are. In contrast to our work, they don't help with guiding the development process to steer the device's security level vs. the resources needed to invest. However, we utilize the CVSS scoring system to prioritize found vulnerabilities enabling measuring the trade-off.

\paragraph{Security Test Suites and Frameworks}
	In order to systematically analyze the security of \ac{IoT} Devices testbeds \cite{Tekeoglu2016}, security test suites \cite{Loi2017}, vulnerability identification processes \cite{Costa2019} and analysis frameworks for smart cameras \cite{Alharbi2018} have been proposed.
	These frameworks should help to identify security and privacy vulnerabilities more systematically based on threat models and experiments.
	The proposed security test suite from Loi et al.~\cite{Loi2017} also helps to determine a three-level, easy-to-read security rating for end customers.
	However, this test suite only focuses on network-based attacks against the communication between the device and the vendor's cloud servers.
	The analysis framework from Alharbi et al.~\cite{Alharbi2018} additionally considers attacks against available web interfaces and communication between the smartphone and cloud servers or the smartphone and the \smarthome{} Device.

Our work systematically aggregates and utilizes methodologies and approaches from previous works on test suites and frameworks \cite{Tekeoglu2016, Loi2017, Costa2019, Alharbi2018}. Additionally, the test cases in this analysis are aggregated from and based on \ac{IoT} security \aclp{BP}, which was not the case or not so systematically done in previous analyses. Our work can be therefore summarized as a systematic security analysis based on \ac{IoT} security \aclp{BP} covering network-based, local configuration network and physical attack vectors and evaluating the \aclp{BP} in the light of found vulnerabilities, which differentiates our work from related work.

\section{Conclusion}
So far, investments in IoT security have been lacking tools for determining what security measures provide the best trade-off in terms of security gains and investment costs. 
In this paper, we introduce a methodology for empirically evaluating the efficacy of IoT Best Practices based on estimating the reduction in potential harm that individual Best Practices affect. We perform an extensive analysis of recent IoT Best Practices proposed by numerous relevant players in the IoT security community and use these as a basis for a comprehensive collection of IoT Best Practices. Based on test cases derived from the Best Practices, we evaluate the security of nine representative real-world IoT devices and use the results of 18 vulnerabilities as an empirical basis for evaluating the efficacy of the gathered Best Practices, providing an indicative ranking for them. We think that the methodology provided in this paper can be adopted by future designers and implementers of IoT devices to guide their investments in security, thus maximizing security gains obtained for the available security budgets.

\clearpage

\bibliographystyle{plain}
\bibliography{bib}
\section*{Appendix}
\appendix
\section{IoT Security Best Practices} \label{sec:app1}
        \paragraph{\acl{HW}}

			\begin{enumerate}[label={BP-HW-\protect\twodigits{\theenumi}}, wide = 0pt]
				\item Use a hardware-based root of trust for, e.g., a secure boot functionality. \cite{EUAFNAIS2017, ISF2019, CultureUKD2018, IEEECommunity2017} \label{BP-HW-Root}
				\item Integrate processors with security chips or Trusted Execution Environments in the device design to prevent attacks against the integrity of code and data. \cite{EUAFNAIS2017, ISF2019, CultureUKD2018, Sharma2019} \label{BP-HW-TEE}
				\item Use pre-provisioned secure elements with private keys which can be used to sign data. The secure element never reveals the private key. \cite{Sharma2019} \label{BP-HW-SecurePK}
				\item Remove or disable test points and debug interfaces or make them physically inaccessible. \cite{EUAFNAIS2017, IEEECommunity2017, ISF2019, Sharma2019} \label{BP-HW-RemoveDebug}
				\item If physical administration interfaces are necessary, they should have effective access protection. \cite{EUAFNAIS2017, IEEECommunity2017, ISF2019} \label{BP-HW-AdminInter}
				\item Protect the \ac{PCB} against physical tampering by using, e.g., epoxy or resin protection. \cite{EUAFNAIS2017} \label{BP-HW-Epoxy}
				\item Make the device's hardware tamper-proof especially when the device is used in public or accessible areas. \cite{EUAFNAIS2017, IEEECommunity2017, ISF2019} \label{BP-HW-TamperProof}
				\item Design the device and the packaging to be tamper-evident, e.g., with tamper-evident seals.~\cite{EUAFNAIS2017, IEEECommunity2017, ISF2019} \label{BP-HW-TamperEvident}
				\item Consider design measures against side-channel attacks for high-security such as masking or shielding. \cite{EUAFNAIS2017, ISF2019, Sharma2019} \label{BP-HW-SideChannel}
				\item Implement a factory-set and tamper-resistant identifier to unambiguously identify a device, e.g., based on secure elements.~\cite{ISF2019} \label{BP-HW-FactorySet}
			\end{enumerate}
\vspace{-0.7em}
		\paragraph{\acf{TIM}}

			\begin{enumerate}[label={BP-TIM-\protect\twodigits{\theenumi}}, wide = 0pt]
				\item A root of trust for secure boot must be established from read-only memory before trusting any other code. There should be no way to bypass this secure boot functionality. Therefore, use a multi-stage bootloader where each stage verifies the next one before execution. Each boot stage additionally verifies the connected hardware components and checks that only the configured ones are present. \cite{EUAFNAIS2017, ISF2019, Sharma2019} \label{BP-TIM-Bootloader}
				\item Sign and verify code in RAM before executing it and/or apply measures to prevent manipulation of code and data after loading it.~\cite{EUAFNAIS2017, ISF2019} \label{BP-TIM-VerifyCode}
				\item Apply measures that prevent the installation of additional possible malicious programs. \cite{EUAFNAIS2017} \label{BP-TIM-Install}
				\item Provide a fail-safe system state which is entered after a security breach or when a failed firmware update occurs. \cite{EUAFNAIS2017, ISF2019} \label{BP-TIM-FailSafe}
				\item Deploy protocols that are able to verify relationships based on trust. \cite{EUAFNAIS2017, Sharma2019} \label{BP-TIM-Relationships}
			\end{enumerate}
\vspace{-0.7em}
		\paragraph{\acf{DSP}}

			\begin{enumerate}[label={BP-DSP-\protect\twodigits{\theenumi}}, wide = 0pt]
				\item Enable all possible security features by default and disable insecure and unnecessary features. \cite{EUAFNAIS2017, ISF2019} \label{BP-Default-Strong}
				\item Use individual default passwords for each device that are hard to crack. \cite{EUAFNAIS2017, CultureUKD2018, IEEECommunity2017} \label{BP-Default-PWCrack}
			\end{enumerate}
\vspace{-0.7em}
		\paragraph{\acf{DPC}}

			\begin{enumerate}[label={BP-DPC-\protect\twodigits{\theenumi}}, wide = 0pt]
				\item Define and document a data classification scheme for all data stored and processed. \cite{ISF2019}  \label{BP-DPC-DataScheme}
				\item Always use informed consent to collect and process personal data.~\cite{EUAFNAIS2017} \label{BP-DPC-Consent}
				\item Only use the collected data for specified and declared purposes.~\cite{EUAFNAIS2017} \label{BP-DPC-CollectedDataPurposes}
				\item Collect only the minimum amount of data necessary to provide a service. \cite{EUAFNAIS2017} \label{BP-DPC-MiniumAmountData}
				\item \ac{IoT} vendors and cloud providers must comply with data protection regulations and laws. \cite{EUAFNAIS2017, CultureUKD2018, ISF2019} \label{BP-DPC-Laws}
				\item Provide means with which the users are able to configure their privacy level and easy personal data deletion. \cite{CultureUKD2018} \label{BP-DPC-DataDeletion}
				\item Resetting a device to factory settings must remove all private data and credentials. \cite{ISF2019} \label{BP-DPC-DataFactoryReset}
			\end{enumerate}
\vspace{-0.7em}
		\paragraph{\acf{SSR}}

			\begin{enumerate}[label={BP-SSR-\protect\twodigits{\theenumi}}, wide = 0pt]
				\item Prevent unacceptable (physical) damage by considering system failures during development. \cite{EUAFNAIS2017} \label{BP-SSR-PhysicalDamage}
				\item Add self-diagnostic and self-repair functions to allow the device to recover on its own. \cite{EUAFNAIS2017} \label{BP-SSR-DiagnosticRepair}
				\item Do not rely on communication channels or cloud services. Enable the system to provide essential functionalities without relying on network services and take power outages into consideration where practical. \cite{EUAFNAIS2017, CultureUKD2018, ISF2019} \label{BP-SSR-RelyNetworkPower}
			\end{enumerate}
\vspace{-0.7em}
		\paragraph{\acf{FU}}

			\begin{enumerate}[label={BP-FU-\protect\twodigits{\theenumi}}, wide = 0pt]
				\item Update the firmware and software of the device at regular intervals: \cite{EUAFNAIS2017, CultureUKD2018, IEEECommunity2017, ISF2019} \label{BP-FU-FWUpdate}
				      \begin{itemize}
					      \item Secure the update infrastructure against attacks.
					      \item Transmit the update files over a secure channel.
					      \item Do not use sensitive information such as hard-coded passwords in firmware update files.
					      \item Sign and encrypt the firmware update file by a trusted authority, e.g., to prevent reverse engineering.
					      \item Verify the firmware update signature on the device before applying it.
					      \item The device should still be able to operate during an update procedure.
				      \end{itemize}
				\item Prevent rollbacks to earlier firmware versions with published security vulnerabilities. \cite{ISF2019}  \label{BP-FU-Rollback}
				\item Firmware updates should be rolled out by an automatic mechanism. \cite{EUAFNAIS2017}  \label{BP-FU-AutomaticFU}
				\item Firmware updates should not change the user's privacy or security settings without informing the user. \cite{EUAFNAIS2017}  \label{BP-FU-Settings}
				\item Consider using an encrypted file system to protect data at rest.~\cite{ISF2019}  \label{BP-FU-EncryptedFS}
			\end{enumerate}
\vspace{-0.7em}
		\paragraph{\acf{AE}}

			\begin{enumerate}[label={BP-AE-\protect\twodigits{\theenumi}}, wide = 0pt]
				\item Implement a device-specific authentication and authorization scheme considering the device's threat model. \cite{EUAFNAIS2017} \label{BP-AUTHEN-DeviceSpecific}
				\item Always change default passwords and usernames during the initial setup. Ensure that only strong passwords can be set by the user.~\cite{EUAFNAIS2017, Payne2017} \label{BP-AUTHEN-StronPW}
				\item Consider implementing two-factor authentication or multi-factor authentication. \cite{EUAFNAIS2017, IEEECommunity2017, ISF2019} \label{BP-AUTHEN-2FA}
				\item Use industry-standard salts, hashes, and if possible, in addition, cryptographic functions to store credentials securely, e.g., on a secure hardware module like Trusted Platform Module or Secure Enclave. \cite{EUAFNAIS2017, CultureUKD2018, ISF2019} \label{BP-AUTHEN-IndustryHashs}
				\item Implement measures to prevent brute-force attacks against login mechanisms. \cite{EUAFNAIS2017} \label{BP-AUTHEN-BruteForce}
				\item Implement a robust password reset mechanism that does not disclose helpful information for an attacker. \cite{EUAFNAIS2017} \label{BP-AUTHEN-PWReset}
				\item Use a trusted and reliable time source to verify certificates.~\cite{ISF2019} \label{BP-AUTHEN-ReliableTimeSource}
			\end{enumerate}
\vspace{-0.7em}
		\paragraph{\acf{AO}}

			\begin{enumerate}[label={BP-AO-\protect\twodigits{\theenumi}}, wide = 0pt]
				\item Implement the principle of least privilege and let services and programs only use the lowest privilege level needed (no root capabilities). \cite{EUAFNAIS2017, CultureUKD2018, ISF2019} \label{BP-AUTHOR-LeastPriviledge}
				\item Do not grant privileges to write to the root file system to users and applications. \cite{ISF2019} \label{BP-AUTHOR-RootFS}
			\end{enumerate}
\vspace{-0.7em}
		\paragraph{\acf{AC}}

			\begin{enumerate}[label={BP-AC-\protect\twodigits{\theenumi}}, wide = 0pt]
				\item Use access control mechanisms to protect the data integrity and confidentiality of sensitive information. \cite{EUAFNAIS2017} \label{BP-ACCESS-Integrity}
				\item Tamper protection, detection, and reaction should not rely on communication channels. \cite{EUAFNAIS2017} \label{BP-ACCESS-TamperComm}
				\item Make it difficult to disassemble the device and encrypt the data storage at rest. \cite{EUAFNAIS2017} \label{BP-ACCESS-Disassemble}
				\item Hardware should only offer ports required to operate the device and debug interfaces should be secured. Use a verification scheme to authenticate valid communication partners on testing interfaces. \cite{EUAFNAIS2017, CultureUKD2018} \label{BP-ACCESS-Ports}
			\end{enumerate}
\vspace{-0.7em}
		\paragraph{\acf{CRYPTO}}

			\begin{enumerate}[label={BP-CRYPTO-\protect\twodigits{\theenumi}}, wide = 0pt]
				\item Use the strongest cryptographic protocols and algorithms whenever possible to protect the confidentiality, authenticity, and integrity of transmitted data and data at rest. Disable weak and insecure protocols. \cite{EUAFNAIS2017, IEEECommunity2017, ISF2019, Payne2017, Sharma2019} \label{BP-CRYPTO-Proto}
				\item Digital certificates and cryptographic keys must be managed securely and there should be an update mechanism for revoking and renewing keys and certificates. \cite{EUAFNAIS2017, ISF2019} \label{BP-CRYPTO-Certs}
				\item Design the cryptographic key management to be scalable. \cite{EUAFNAIS2017} \label{BP-CRYPTO-Scalable}
			\end{enumerate}
\vspace{-0.7em}
		\paragraph{\acf{STC}}

			\begin{enumerate}[label={BP-STC-\protect\twodigits{\theenumi}}, wide = 0pt]
				\item Protect information on the \ac{IoT} Device, in transit and the cloud to ensure confidentiality, integrity, availability, and authenticity.~\cite{EUAFNAIS2017, IEEECommunity2017} \label{BP-STC-Transit}
				\item Use open, standardized, and the most recent version of security protocols, e.g., TLS/SSL. \cite{EUAFNAIS2017, CultureUKD2018, IEEECommunity2017} \label{BP-STC-SecurityProto}
				\item Protect credentials and sensitive data in transit on internal and external networks. Never transmit credentials in clear text over wireless networks. \cite{EUAFNAIS2017, CultureUKD2018, ISF2019} \label{BP-STC-Credentials}
				\item Always sign data cryptographically. \cite{EUAFNAIS2017} \label{BP-STC-Sign}
				\item Received data should be verified first and connections to other devices and services should be established only after verifying the identity. \cite{EUAFNAIS2017} \label{BP-STC-Verified}
				\item Disable unnecessary ports and services, only include services required for the device's functionality. \cite{EUAFNAIS2017, CultureUKD2018, ISF2019, Payne2017} \label{BP-STC-PortsDisable}
				\item Ship the latest stable version of the operating system on the device.~\cite{ISF2019} \label{BP-STC-OSShipment}
				\item Limit the amount of bandwidth sent and received to prevent \ac{DDoS} attacks originating from the device. \cite{EUAFNAIS2017, IEEECommunity2017} \label{BP-STC-LimitBandwidth}
			\end{enumerate}
\vspace{-0.7em}
		\paragraph{\acf{SIS}}

			\begin{enumerate}[label={BP-SIS-\protect\twodigits{\theenumi}}, wide = 0pt]
				\item Organize network elements into different segments to isolate services.  \cite{EUAFNAIS2017, IEEECommunity2017, Payne2017} \label{BP-SINS-Segments}
				\item Compromising a single device should not allow taking over other devices managed by the same protocol. \cite{EUAFNAIS2017} \label{BP-SINS-CompromisingProtocol} 
				\item Do not use the same secret key in the entire product family.~\cite{EUAFNAIS2017} \label{BP-SINS-SecretKey}
				\item Enable only necessary ports and network interfaces. \cite{EUAFNAIS2017, IEEECommunity2017, ISF2019} \label{BP-SINS-NetworkInterfaces}
				\item If possible use a properly configured firewall. \cite{ISF2019, Payne2017} \label{BP-SINS-Firewall}
				\item Make the infrastructure and device \ac{DDoS}-resilient and enable the infrastructure to load balance traffic. \cite{EUAFNAIS2017, CultureUKD2018} \label{BP-SINS-DDoS}
				\item Encrypt the user session on web interfaces and also on the way to the backend. Protect the interface against common web security vulnerabilities like XSS, CSRF, and SQL injection. \cite{EUAFNAIS2017} \label{BP-SINS-UserSession}
			\end{enumerate}
\vspace{-0.7em}
		\paragraph{\acf{SIOH}}

			\begin{enumerate}[label={BP-SIOH-\protect\twodigits{\theenumi}}, wide = 0pt]
				\item Validate data inputs from user interfaces and other network services and filter outgoing data.  \cite{EUAFNAIS2017, CultureUKD2018, ISF2019} \label{BP-SIOH-Filter}
			\end{enumerate}
\vspace{-0.7em}
		\paragraph{\acf{LOG}}

			\begin{enumerate}[label={BP-LOG-\protect\twodigits{\theenumi}}, wide = 0pt]
				\item Use a logging system to document security events like user authentication. Log files should be accessible to authenticated parties only and saved on non-volatile storage for security anomaly scans. The user should be informed about the type of data being collected and the reason for collecting it. \cite{EUAFNAIS2017, CultureUKD2018} \label{BP-LOG-LoggingSystem}
				\item Store logs on a different system partition and run the logging service on its own system process. Rotate logs and set a maximum log size. \cite{ISF2019} \label{BP-LOG-OwnProcess}
				\item Do not log any sensitive private information and credentials.~\cite{ISF2019} \label{BP-LOG-Credentials}
			\end{enumerate}
\vspace{-0.7em}
		\paragraph{\acf{MAT}}

			\begin{enumerate}[label={BP-MAT-\protect\twodigits{\theenumi}}, wide = 0pt]
				\item Monitor the device's behavior on a regular basis to be able to detect malware and integrity errors. Unauthorized changes should trigger an alert to inform the user or administrator about the issue. \cite{EUAFNAIS2017, CultureUKD2018} \label{BP-MAT-MonitorMalware}
				\item Perform periodic audits and security measure reviews. These should determine whether the chosen security measures are still effective. Penetration testing should be done at least semiannual.  \cite{EUAFNAIS2017} \label{BP-MAT-AuditsMeasures}
				\item Perform dynamic testing in addition to static testing to find vulnerabilities in commodity hardware. \cite{IEEECommunity2017} \label{BP-MAT-DynamicTesting}
			\end{enumerate}

\section{Test Cases} \label{sec:app3}
\paragraph{\acf{HW}}\label{sec:TestCaseSelection:Hardware}
\begin{enumerate}[label={TC-HW-\protect\twodigits{\theenumi}}, wide = 0pt]
    \item Checks whether the device uses a hardware-based root of trust, a security chip, a Trusted Execution Environment, or pre-provisioned secure elements to verify data. (\ref{BP-HW-Root}, \ref{BP-HW-TEE}, \ref{BP-HW-SecurePK}, \ref{BP-TIM-Bootloader} and \ref{BP-TIM-VerifyCode}) \label{TC-HW-SecureChip}
    \item Checks whether measures against physical tampering have been applied, e.g., epoxy or resin protection for the \ac{PCB} or whether the device is hard to disassemble. (\ref{BP-HW-Epoxy}, \ref{BP-HW-TamperProof} and \ref{BP-ACCESS-Disassemble}) \label{TC-HW-TamperCase}
    \item Check whether the package uses a tamper-evident seal. (\ref{BP-HW-TamperEvident}) \label{TC-HW-Seal}
    \item Checks if there are any helpful test points, debug interfaces, or administrative ports like \ac{UART} interfaces and if they are accessible. (\ref{BP-HW-RemoveDebug}, \ref{BP-HW-AdminInter} and \ref{BP-ACCESS-Ports}) \label{TC-HW-Ports}
\end{enumerate}\vspace{-0.7em}
\paragraph{\acf{TIM}}\label{sec:TestCaseSelection:Trust}
\begin{enumerate}[label={TC-TIM-\protect\twodigits{\theenumi}}, wide = 0pt]
    \item Check if it is possible to install additional software or scripts, e.g., via an administrative interface. (\ref{BP-TIM-Install}) \label{TC-TIM-Software}
\end{enumerate}\vspace{-0.7em}
\paragraph{\acf{DSP}}\label{sec:TestCaseSelection:DefaultSecurity}
\begin{enumerate}[label={TC-DSP-\protect\twodigits{\theenumi}}, wide = 0pt]
    \item Checks whether the device uses strong and individual default credentials. (\ref{BP-Default-PWCrack}) \label{TC-DSP-DefaultSecurity}
\end{enumerate}\vspace{-0.7em}
\paragraph{\acf{DPC}}\label{sec:TestCaseSelection:Protection}
\begin{enumerate}[label={TC-DPC-\protect\twodigits{\theenumi}}, wide = 0pt]
    \item Checks whether private data or credentials like \wifi{} passwords are removed by a factory reset. This can be checked, e.g., by resetting the device and reading out the flash memory with an external flash reader afterward. (\ref{BP-DPC-DataFactoryReset}) \label{TC-DPC-DataFactoryReset}
\end{enumerate}\vspace{-0.7em}
\paragraph{\acf{SSR}}\label{sec:TestCaseSelection:SystemSafety}
\begin{enumerate}[label={TC-SSR-\protect\twodigits{\theenumi}}, wide = 0pt]
    \item It is tested whether the device relies on an internet connection or cloud services to function properly and still provides basic functionalities during an internet connection loss. (\ref{BP-SSR-RelyNetworkPower}) \label{TC-SSR-RelyNetwork}
\end{enumerate}\vspace{-0.7em}
\paragraph{\acf{FU}}\label{sec:TestCaseSelection:Firmware}
\begin{enumerate}[label={TC-FU-\protect\twodigits{\theenumi}}, wide = 0pt]
    \item Tests whether the device uses an automatic firmware update mechanism. (\ref{BP-FU-FWUpdate} and \ref{BP-FU-AutomaticFU}) \label{TC-FU-UpdateFU}
    \item Checks whether the update is transmitted over a secure channel. (\ref{BP-FU-FWUpdate}) \label{TC-FU-SecureFU}
    \item Checks whether the firmware update contains any hard-coded passwords or credentials. (\ref{BP-FU-FWUpdate}) \label{TC-FU-Passwords}
    \item Verifies whether the firmware update is encrypted and signed. (\ref{BP-FU-FWUpdate}) \label{TC-FU-Encrypt}
    \item Checks whether the (signed) firmware update is verified before executing it. (\ref{BP-FU-FWUpdate}) \label{TC-FU-Verify}
    \item Checks whether rollbacks to earlier firmware versions are possible. (\ref{BP-FU-Rollback}) \label{TC-FU-Rollback}
    \item Checks whether an encrypted file system is used to protect the data at rest. (\ref{BP-FU-EncryptedFS} and \ref{BP-CRYPTO-Proto}) \label{TC-FU-EncryptedFS}
\end{enumerate}\vspace{-0.7em}
\paragraph{\acf{AE}}\label{sec:TestCaseSelection:Authentication}
\begin{enumerate}[label={TC-AE-\protect\twodigits{\theenumi}}, wide = 0pt]
    \item Check whether the device's default credentials are changed during the initial setup and only strong passwords can be set. (\ref{BP-AUTHEN-StronPW}) \label{TC-AE-StrongPW}
    \item Verify that the app or the device provides a two-factor or multi-factor authentication method. (\ref{BP-AUTHEN-2FA}) \label{TC-AE-2FA}
    \item Verify that the device's credentials are stored securely on the device via industry-standard hash functions. (\ref{BP-AUTHEN-IndustryHashs} and \ref{BP-CRYPTO-Proto}) \label{TC-AE-IndustryHash}
    \item Check whether login websites or shell logins implement measures to prevent brute-force attacks. (\ref{BP-AUTHEN-BruteForce}) \label{TC-AE-BruteForce}
\end{enumerate}\vspace{-0.7em}
\paragraph{\acf{AO}}\label{sec:TestCaseSelection:Authorization}
\begin{enumerate}[label={TC-AO-\protect\twodigits{\theenumi}}, wide = 0pt]
    \item Verify that services only have the absolute minimum privilege level and do not use root capabilities. (\ref{BP-AUTHOR-LeastPriviledge}) \label{TC-AO-LeastPriviledge}
    \item Check whether normal users can write to the root file system. (\ref{BP-AUTHOR-RootFS}) \label{TC-AO-RootFS}
\end{enumerate}\vspace{-0.7em}
\paragraph{\acf{AC}}\label{sec:TestCaseSelection:AccessControl}
The device's disassembling process (\ref{BP-ACCESS-Disassemble}) is tested by the test case \ref{TC-HW-TamperCase} and the physical ports (\ref{BP-ACCESS-Ports}) are tested by \ref{TC-HW-Ports}.
\paragraph{\acf{CRYPTO}}\label{sec:TestCaseSelection:Crypto}
Already covered by existing test cases (\ref{BP-CRYPTO-Proto}).
\paragraph{\acf{STC}}\label{sec:TestCaseSelection:Communication}
\begin{enumerate}[label={TC-STC-\protect\twodigits{\theenumi}}, wide = 0pt]
    \item Check whether data are properly protected during transmission (e.g., in \wifi{}, Bluetooth or Ethernet connections) by standardized protocols like TLS/SSL and credentials are not transmitted in plain text. (\ref{BP-STC-Transit}, \ref{BP-STC-SecurityProto}, \ref{BP-STC-Credentials} and \ref{BP-CRYPTO-Proto}) \label{TC-STC-SecureTransit}
    \item Check whether transmitted data are additionally signed. (\ref{BP-STC-Sign} and \ref{BP-CRYPTO-Proto}) \label{TC-STC-Singed}
    \item Check that communication endpoints are verified first before performing a data exchange. (\ref{BP-STC-Verified},  \ref{BP-TIM-Relationships} and \ref{BP-CRYPTO-Proto}) \label{TC-STC-Verified}
    \item Check if there are (unnecessarily) open ports and services running on the device. (\ref{BP-Default-Strong} and \ref{BP-STC-PortsDisable}) \label{TC-STC-Ports}
    \item Analyze the device's operability during \ac{DoS} attacks and whether there are, e.g., firewalls or kernel configurations that restrict the incoming and outgoing traffic. (\ref{BP-STC-LimitBandwidth}, \ref{BP-SINS-DDoS}) \label{TC-STC-LimitBandwidth}
\end{enumerate}\vspace{-0.7em}
\paragraph{\acf{SIS}}\label{sec:TestCaseSelection:Interfaces}
\begin{enumerate}[label={TC-SIS-\protect\twodigits{\theenumi}}, wide = 0pt]
    \item Check whether the device uses a firewall. (\ref{BP-SINS-Firewall}) \label{TC-SIS-Firewall}
    \item If there are any web interfaces, verify that they are properly protected against common web security vulnerabilities like XSS, CSRF, SQL injection, and unauthenticated services. (\ref{BP-SINS-UserSession} and \ref{BP-SINS-NetworkInterfaces}) \label{TC-SIS-WebVulnerabilities}
    \item Check that only necessary network interfaces are enabled. (\ref{BP-SINS-NetworkInterfaces}) \label{TC-SIS-NetworkInterfaces}
\end{enumerate}\vspace{-0.7em}
\paragraph{\acf{SIOH}}\label{sec:TestCaseSelection:InputOutput}
\begin{enumerate}[label={TC-SIOH-\protect\twodigits{\theenumi}}, wide = 0pt]
    \item Test whether it is possible to manipulate user inputs or traffic generated from other services (e.g., a mobile app) to gain unauthorized access to the device, e.g., by exploiting improper data validation and sanitization. (\ref{BP-SIOH-Filter}) \label{TC-SIOH-Filter}
\end{enumerate}\vspace{-0.7em}
\paragraph{\acf{LOG}}\label{sec:TestCaseSelection:Loggin}
\begin{enumerate}[label={TC-LOG-\protect\twodigits{\theenumi}}, wide = 0pt]
    \item  Check if the device uses a logging system and that this logging system has special protection. Additionally, the log should be searched for any sensitive private information or credentials. (\ref{BP-LOG-Credentials}, \ref{BP-LOG-LoggingSystem} and \ref{BP-LOG-OwnProcess}) \label{TC-LOG-Filter}
\end{enumerate}\vspace{-0.7em}
\paragraph{\acf{MAT}}\label{sec:TestCaseSelection:Monitoring}
\begin{enumerate}[label={TC-MAT-\protect\twodigits{\theenumi}}, wide = 0pt]
    \item Investigate whether there are user alerts after successful unauthorized access to a device. (\ref{BP-MAT-MonitorMalware}) \label{TC-MAT-Inform}
\end{enumerate}

\subsection{Best Practice Selection}\label{sec:bps}
    The adherence to some Best Practices cannot be tested by anyone except the manufacturer or Best Practices are not applicable to consumer IoT devices.

    Data protection and compliance recommendations \ref{BP-DPC-DataScheme} -- \ref{BP-DPC-DataDeletion} are relying on vendor-specific internal information and contracts. \ref{BP-AUTHEN-DeviceSpecific} is not covered by a test case because without internal knowledge it is not possible to determine if the vendor did implement the current authentication based on the device's threat model. Password reset mechanisms (\ref{BP-AUTHEN-PWReset}) are normally part of the vendor's cloud and not device-specific. The scalability of key management systems cannot be fully investigated because they are mainly based on cloud components out of reach (\ref{BP-CRYPTO-Certs} and \ref{BP-CRYPTO-Scalable}). The shipped version of the operating system (\ref{BP-STC-OSShipment}) cannot be analyzed as the version reaching the end customer depends highly on the storage time at intermediate warehouses and sellers. Most of the \aclp{BP} from the subcategory \enquote{Secure Interfaces and network services} cannot be investigated because they rely on specific knowledge about the cloud or the whole product family (\ref{BP-SINS-Segments} -- \ref{BP-SINS-SecretKey}).

    Similarly, a test case for a fail-safe system state is not defined (\ref{BP-TIM-FailSafe}) as the tested devices are not responsible for vital systems. Additionally, possible direct (physical) damage (\ref{BP-SSR-PhysicalDamage}) and self-repair functionality (\ref{BP-SSR-DiagnosticRepair}) are not applicable to consumer IoT devices. Possible silent changes to privacy settings caused by firmware updates are not tested because of the low relevance for the device security (\ref{BP-FU-Settings}). The usage of a trusted and reliable time source is excluded (\ref{BP-AUTHEN-ReliableTimeSource}) as if certificates are used correctly and there is no leakage of private keys, it is not possible to take advantage of an incorrect time source. \ref{BP-ACCESS-Integrity} is excluded as it targets physical access control measures in companies or organizations.

\section{Evaluation} \label{sec:app2}

\begin{table}[h!]
			\myfloatalign
			\caption{Occurrences and average \ac{CVSS} Scores of \aclp{BP} leading to Vulnerabilities}
			\label{tab:AnalysisBPCausingVuln}
            \begin{scriptsize}
			\begin{tabularx}{\linewidth}{Xcc}
				\toprule
				\tableheadline{Best Practice}   & \tableheadline{Occurrence} & \tableheadline{Average CVSS Score}
				\\ \midrule
				\ref{BP-Default-PWCrack}        & 7                          & 6.5                                \\ 
				\ref{BP-CRYPTO-Proto}           & 4                          & 5.1                                \\ 
				\ref{BP-HW-RemoveDebug}         & 3                          & \underline{7.1}                    \\ 
				\ref{BP-HW-AdminInter}          & 3                          & \underline{7.1}                    \\ 
				\ref{BP-LOG-Credentials}        & 2                          & 3.1                                \\ 
				\ref{BP-SIOH-Filter}            & 2                          & \underline{8.8}                    \\ 
				\ref{BP-SINS-UserSession}       & 2                          & 5.3                                \\ 
				\ref{BP-STC-Transit}            & 2                          & 6.6                                \\ 
				\ref{BP-STC-SecurityProto}      & 2                          & 6.6                                \\ 
				\ref{BP-STC-Credentials}        & 2                          & 6.6                                \\ 
				\ref{BP-STC-Verified}           & 2                          & 3.7                                \\ 
				\ref{BP-TIM-Relationships}      & 2                          & 3.7                                \\ 
				\ref{BP-AUTHEN-PWReset}         & 1                          & 5.9                                \\ 
				\ref{BP-SINS-NetworkInterfaces} & 1                          & \underline{8.3}                    \\ 
				\ref{BP-SINS-DDoS}              & 1                          & 5.7                                \\ 
				\bottomrule
			\end{tabularx}
            \end{scriptsize}
			\\
			\vspace{0.5\baselineskip}			
			\begin{small}
			The \underline{underlined values} are high ($\geq$ 7.0) average \ac{CVSS} scores.
			\end{small}
   \vspace{-2em}
		\end{table}

\begin{sidewaystable}[htbp]
            \vskip 0.4\textwidth
			\myfloatalign
			\caption{Potential Security Issues and Vulnerabilities Part 1}\label{tab:AnalysisSummary1}
			\begin{footnotesize}
				\begin{tabularx}{\linewidth}{p{1.4cm}XXXX}
					\toprule
					\tableheadline{Device}                                          & \tableheadline{Information Gathering} & \tableheadline{Nearby Attacker} & \tableheadline{Same Network Attacker} & \tableheadline{Physical Attacker} \\
					\midrule
					Arlo\hyperlink{tab:results:arlo}{\textsuperscript{1}}           & ---                                   & ---                             & \begin{smallitemize}
						\item Onboarding takeover with different user possible
						(\ac{CVSS} \href{https://nvd.nist.gov/vuln-metrics/cvss/v3-calculator?vector=AV:A/AC:H/PR:N/UI:R/S:U/C:L/I:N/A:L\&version=3.1}{3.7})
					\end{smallitemize}        & \begin{smallitemize}
						\item Read-Only \ac{UART} interface
						\item Logging of sensitive data, e.g., access tokens
						(\ac{CVSS} \href{https://nvd.nist.gov/vuln-metrics/cvss/v3-calculator?vector=AV:P/AC:H/PR:N/UI:N/S:U/C:L/I:N/A:N\&version=3.1}{2.0})
					\end{smallitemize}    \\ \hline
					Blink\hyperlink{tab:results:blink}{\textsuperscript{2}}         & ---                                   & \begin{smallitemize}
						\item Unencrypted \wifi{} \& HTTP traffic
						\item Central master key for onboarding encryption
						\item \ac{DoS} attacks possible
						(\ac{CVSS} \href{https://nvd.nist.gov/vuln-metrics/cvss/v3-calculator?vector=AV:A/AC:L/PR:N/UI:R/S:U/C:N/I:N/A:H\&version=3.1}{5.7})
					\end{smallitemize}  & \begin{smallitemize}
						\item Missing SSL certificate verification leads to \ac{MitM} attack vectors
						(\ac{CVSS}~\href{https://nvd.nist.gov/vuln-metrics/cvss/v3-calculator?vector=AV:N/AC:H/PR:N/UI:N/S:U/C:H/I:N/A:H\&version=3.1}{7.4})
					\end{smallitemize}        & ---                               \\ \hline
					Bosch\hyperlink{tab:results:bosch}{\textsuperscript{3}}         & ---                                   & ---                             & \begin{smallitemize}
						\item Missing two-factor authentication
					\end{smallitemize}        & ---                               \\ \hline
					D-Link\hyperlink{tab:results:dlink}{\textsuperscript{4}}        & \begin{smallitemize}
						\item Hardcoded and only weakly hashed credentials in the \OSC{}
					\end{smallitemize}        & \begin{smallitemize}
						\item Insecure \bluetooth{} pairing method (\textit{Just Works} method)
						\item Weak \bluetooth{} authentication scheme via 6-digit PIN
						\item \bluetooth{} is not disabled
						\item Command injection vulnerabilities in \bluetooth{} services
						(\ac{CVSS} \href{https://nvd.nist.gov/vuln-metrics/cvss/v3-calculator?vector=AV:A/AC:H/PR:N/UI:N/S:C/C:H/I:H/A:H\&version=3.1}{8.3})
					\end{smallitemize}  & \begin{smallitemize}
						\item Weak unchangeable \ac{RTSP} service password
						(\ac{CVSS} \href{https://nvd.nist.gov/vuln-metrics/cvss/v3-calculator?vector=AV:A/AC:L/PR:N/UI:N/S:U/C:H/I:N/A:N\&version=3.1}{6.5})
						\item Unauthenticated \ac{RTSP} access via \lstinline[basicstyle=\scriptsize]|StreamProxy| binary
						(\ac{CVSS} \href{https://nvd.nist.gov/vuln-metrics/cvss/v3-calculator?vector=AV:A/AC:H/PR:N/UI:N/S:U/C:H/I:N/A:N\&version=3.1}{5.3})
						\item Weak \& static firmware update encryption key
						\item Hardcoded credentials in the firmware update

					\end{smallitemize}        & \begin{smallitemize}
						\item Unauthenticated \ac{UART} interface
						(\ac{CVSS} \href{https://nvd.nist.gov/vuln-metrics/cvss/v3-calculator?vector=AV:P/AC:H/PR:N/UI:N/S:C/C:H/I:H/A:H\&version=3.1}{7.1})

						\item \telnet{} activation via MicroSD Card file
						(\ac{CVSS} \href{https://nvd.nist.gov/vuln-metrics/cvss/v3-calculator?vector=AV:P/AC:H/PR:N/UI:N/S:C/C:H/I:H/A:H\&version=3.1}{7.1})
						\item Logging of sensitive data, e.g., on \ac{UART} interface
						(\ac{CVSS} \href{https://nvd.nist.gov/vuln-metrics/cvss/v3-calculator?vector=AV:P/AC:H/PR:N/UI:N/S:U/C:H/I:N/A:N\&version=3.1}{4.2})
						\item Missing secure boot functionality
					\end{smallitemize}    \\ \hline
					Nest Camera\hyperlink{tab:results:nestcam}{\textsuperscript{5}} & ---                                   & \begin{smallitemize}
						\item Always enabled \ac{BLE} interface
					\end{smallitemize}  & ---                                   & \begin{smallitemize}
						\item Protected \ac{UART} interface
					\end{smallitemize}
					\\
					\bottomrule
				\end{tabularx}
				\\
				\vspace{\baselineskip}		
				\hypertarget{tab:results:arlo}{\textsuperscript{1}\arlocam{}} \quad \hypertarget{tab:results:blink}{\textsuperscript{2}\blinkcam{}} \quad \hypertarget{tab:results:bosch}{\textsuperscript{3}\boschcam{}} \quad \hypertarget{tab:results:dlink}{\textsuperscript{4}\dlinkcam{}} \quad \hypertarget{tab:results:nestcam}{\textsuperscript{5}\nestcam{}}
			\end{footnotesize}
		\end{sidewaystable}

		\begin{sidewaystable}[htbp]
            \vskip 0.4\textwidth
			\myfloatalign
			\caption{Potential Security Issues and Vulnerabilities Part 2}
			\label{tab:AnalysisSummary2}
			\begin{footnotesize}
				\begin{tabularx}{\linewidth}{p{1.4cm}XXXX}
					\toprule
					\tableheadline{Device}                                                    & \tableheadline{Information Gathering} & \tableheadline{Nearby Attacker} & \tableheadline{Same Network Attacker} & \tableheadline{Physical Attacker} \\
					\midrule

					Nest\quad Protect\hyperlink{tab:results:nestprotect}{\textsuperscript{1}} & ---                                   & \begin{smallitemize}
						\item Reconfiguration via \ac{BLE} interface possible

					\end{smallitemize}  & ---                                   & ---                              
					\\\hline
					Ring\hyperlink{tab:results:ring}{\textsuperscript{2}}                     & ---                                   & \begin{smallitemize}
						\item Unencrypted onboarding \wifi{} network
						\item Lack of onboarding service authentication measures
						(\ac{CVSS} \href{https://nvd.nist.gov/vuln-metrics/cvss/v3-calculator?vector=AV:A/AC:H/PR:N/UI:R/S:U/C:L/I:N/A:L\&version=3.1}{3.7})
					\end{smallitemize}  & ---                                   & ---                                                              
					\\ \hline
					Tesvor\hyperlink{tab:results:tesvor}{\textsuperscript{3}}                 & \begin{smallitemize}
						\item User existence check
						\item Weak password recovery process with 4-digit PIN
						(\ac{CVSS} \href{https://nvd.nist.gov/vuln-metrics/cvss/v3-calculator?vector=AV:N/AC:H/PR:N/UI:N/S:U/C:H/I:N/A:N\&version=3.1}{5.9})
						\item Missing two-factor authentication
					\end{smallitemize}        &
					\begin{smallitemize}
						\item Unencrypted \wifi{} with no authentication
						\item Leakage of \wifi{} credentials during onboarding
						(\ac{CVSS} \href{https://nvd.nist.gov/vuln-metrics/cvss/v3-calculator?vector=AV:A/AC:L/PR:N/UI:R/S:U/C:H/I:N/A:N\&version=3.1}{5.7})
					\end{smallitemize}
					                                                                          & \begin{smallitemize}

						\item Missing automatic firmware update mechanism
						\item Protected SSH service
					\end{smallitemize}        & \begin{smallitemize}
						\item Disabled \ac{UART} interface
					\end{smallitemize}                                                                              \\ \hline
					TP-Link\hyperlink{tab:results:tplink}{\textsuperscript{4}}                & \begin{smallitemize}
						\item Hardcoded credentials in \OSC{}
					\end{smallitemize}        & \begin{smallitemize}
						\item Unencrypted onboarding \wifi{} network
						\item \ac{RTSP} authentication bypass in \lstinline[basicstyle=\scriptsize]|cet| binary
						(\ac{CVSS} \href{https://nvd.nist.gov/vuln-metrics/cvss/v3-calculator?vector=AV:A/AC:H/PR:N/UI:N/S:U/C:H/I:N/A:N\&version=3.1}{5.3})

						\item Command injection vulnerability in \lstinline[basicstyle=\scriptsize]|uhttpd| binary
						(\ac{CVSS} \href{https://nvd.nist.gov/vuln-metrics/cvss/v3-calculator?vector=AV:A/AC:L/PR:N/UI:N/S:C/C:H/I:H/A:H\&version=3.1}{9.3})
					\end{smallitemize}  & \begin{smallitemize}
						\item Missing two-factor authentication
						\item \ac{RTSP} authentication bypass in \lstinline[basicstyle=\scriptsize]|cet| binary
						(\ac{CVSS} \href{https://nvd.nist.gov/vuln-metrics/cvss/v3-calculator?vector=AV:A/AC:H/PR:N/UI:N/S:U/C:H/I:N/A:N\&version=3.1}{5.3})
						\item Command injection vulnerability in \lstinline[basicstyle=\scriptsize]|uhttpd| binary
						(\ac{CVSS} \href{https://nvd.nist.gov/vuln-metrics/cvss/v3-calculator?vector=AV:A/AC:L/PR:N/UI:N/S:C/C:H/I:H/A:H\&version=3.1}{9.3})
						\item Missing automatic firmware update mechanism
						\item Hardcoded password in the firmware update
						\item Firmware downgrade attack possible
					\end{smallitemize}        & \begin{smallitemize}
						\item Improperly protected \ac{UART} interface with root shell
						(\ac{CVSS} \href{https://nvd.nist.gov/vuln-metrics/cvss/v3-calculator?vector=AV:P/AC:H/PR:N/UI:N/S:C/C:H/I:H/A:H\&version=3.1}{7.1})
						\item Missing secure boot functionality
					\end{smallitemize}    \\
					\bottomrule
				\end{tabularx}
				\\		
				\vspace*{\baselineskip}
				\hypertarget{tab:results:nestprotect}{\textsuperscript{1}\nestprotect{}} \quad \hypertarget{tab:results:ring}{\textsuperscript{2}\ringcam{}} \quad \hypertarget{tab:results:tesvor}{\textsuperscript{3}\tesvor{}} \quad \hypertarget{tab:results:tplink}{\textsuperscript{4}\tpcam{}}
			\end{footnotesize}
		\end{sidewaystable}

\end{document}